\documentclass[journal]{IEEEtran}

\usepackage{amsmath}
\usepackage{amssymb}
\usepackage{amsfonts}
\usepackage{float}
\usepackage{algorithm}
\usepackage{algpseudocode}
\usepackage{booktabs}
\usepackage{siunitx}  
\usepackage{multirow}
\usepackage{dashrule}
\usepackage{graphicx}

\IEEEoverridecommandlockouts{}                         

\title{\LARGE \bf An integrated interpretable control effectiveness learning and nonlinear control allocation methodology for overactuated aircraft}

\author{Umut Demir$^{1}$, Aamir Ahmad$^{1}$, Torbj\o rn Cunis$^{1}$ and  Walter Fichter$^{1}$
\thanks{$^{1}$University of Stuttgart, Faculty of Aerospace Engineering and Geodesy, Institute of Flight Mechanics and Control (iFR), Germany.
        {\tt\small firstname.lastname@ifr.uni-stuttgart.de}}}

\begin{document}

\maketitle
\thispagestyle{empty}
\pagestyle{empty}

\begin{abstract}
Nonlinear dynamics and the strong couplings that arise between multiple effectors 
undermine the assumptions behind conventional, linear control allocation techniques. 
When flight enters regimes where nonlinear effects dominate, linear allocators 
exhibit reduced accuracy due to increased model mismatch, which subsequently 
degrades performance and robustness of the flight control system. High fidelity 
onboard models and black box data driven approaches can recover accuracy across 
the flight envelope, but respectively impose computational burdens prohibitive for 
real time allocation and sacrifice the interpretability required for verification 
and fault diagnosis. This paper addresses these limitations by learning an explicit, physics constrained analytical model of the control effectiveness mapping from
representative flight data using Sparse Identification of Nonlinear Dynamics. The resulting mapping is  compact, interpretable, and admits analytical derivatives, enabling efficient computation within nonlinear solvers that additionally incorporate actuator dynamics, without requiring an onboard model. An online adaptation mechanism monitors prediction residuals 
and refreshes the model when significant plant changes are detected, providing 
graceful reconfiguration under actuator failures and varying operating conditions. 
The methodology is evaluated on a high fidelity nonlinear benchmark aircraft across 
a range of aggressive maneuvers, achieving 
accuracy comparable to a full nonlinear onboard model while substantially reducing 
computational cost relative to established baselines.
\end{abstract}


\section{INTRODUCTION}\label{sec:introduction}
Modern engineered systems, from autonomous robots with several effectors to 
satellites equipped with clusters of reaction wheels, are increasingly 
overactuated~\cite{yin2016review}~\cite{4124854}. They possess more effectors 
than independent forces or moments that must be produced. Control allocation is 
the layer that converts a high level command, such as a desired torque, thrust 
vector, or generalized virtual input, into a specific set of actuator commands 
while respecting physical limits. Beyond satisfying this primary tracking 
objective, the inherent redundancy in the actuator set provides additional degrees 
of freedom that can be exploited to satisfy secondary objectives such as minimizing 
power usage, limiting surface deflection rates, reducing noise, or preserving a 
low radar infrared signature for stealth~\cite{yang2019new}. This redundancy is purposeful by design: a larger actuator set improves fault 
tolerance and reconfiguration capability, and is often preferred for reasons of 
safety, accuracy, bandwidth, and maintainability. In complex platforms, effector 
groups may serve multiple control objectives simultaneously; for example, in 
aircraft, engines are primarily designed for thrust yet can be commanded to 
generate a yawing moment when conventional aerodynamic surfaces are insufficient 
or unavailable~\cite{durham2017aircraft}~\cite{CAsurvey}.

\begin{figure}[ht]
    \centering
    \includegraphics[width=0.99\linewidth]{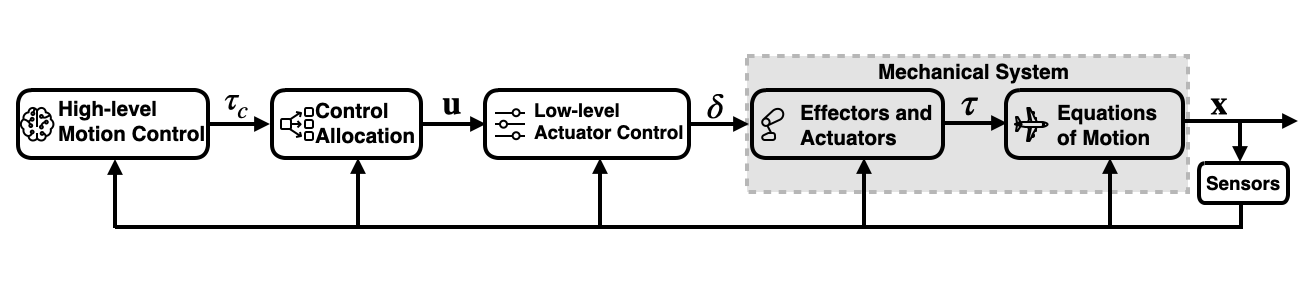}
    \caption{Control system architecture with control allocation. The high level 
    motion controller commands a virtual effort vector $\tau_c$; the allocation 
    layer distributes this demand across individual effectors}\label{fig:highlevelCA}
\end{figure}

As illustrated in Fig.~\ref{fig:highlevelCA}, the control allocation layer
distributes the virtual control commanded by the flight controller to actuator
commands. In general, the mapping from actuator commands to the resulting
virtual control is a nonlinear function of the vehicle state and the actuator
commands,
\begin{equation}
  \boldsymbol{\tau_{c}} = \mathbf{f}(\mathbf{x},\,\mathbf{u}),
  \label{eq:nonlinearmodel}
\end{equation}
which captures nonlinear dynamics and state dependent variations in control effectiveness. In the standard
formulation of control allocation, this mapping is approximated by a
linear model~\cite{oppenheimer2006control},
\begin{equation}
  \boldsymbol{\tau_{c}} = \mathbf{B}\,\mathbf{u},
  \label{eq:linearmodel}
\end{equation}
where $\mathbf{B}\in\mathbb{R}^{n\times m}$ is the control effectiveness
matrix, $\boldsymbol{\tau_{c}}\in\mathbb{R}^{n}$ is the virtual control vector,
and $\mathbf{u}\in\mathbb{R}^{m}$ is the actuator command vector. Each entry
$B_{ij}=\partial\tau_i/\partial u_j$ represents the sensitivity of the
$i$\textsuperscript{th} virtual control to the $j$\textsuperscript{th}
effector. In control allocation problems, effector redundancy implies $m>n$, so $\mathbf{B}$ is rectangular~\cite{oppenheimer2006control}.

The literature on solution methods for linear control allocation is extensive. A number of well established approaches are widely cited in the literature. Explicit ganging methods precombine multiple effectors so that multiple devices act as a single, composite control input determined a priori~\cite{oppenheimer2006control}. The pseudoinverse approach is an optimization based method that relies on computing the pseudoinverse of the typically nonsquare effectiveness matrix~\cite{durham2017aircraft}. Because solutions are non unique, null space methods augment any particular solution with components in the null space of the effectiveness matrix to satisfy secondary objectives without altering the achieved generalized forces and moments, or simple saturations, heuristic schemes such as redistributed pseudoinverse and daisy chaining iteratively freeze saturated actuators and reallocate the remaining demand over the unsaturated set, repeating this reduced problem until feasibility or no further improvement is obtained~\cite{daisy}. 

Linear control allocation has intrinsic drawbacks since it freezes a time varying, state dependent mapping into a static and linear effectiveness matrix. In practice, the true effectiveness of the control often changes with operating conditions and the effector state. For aircrafts, aerodynamic surfaces vary with airspeed and angle of attack and exhibit geometric nonlinearities; flow conditions depend on both vehicle and effector states; and interactions arise among effectors and between effectors and the body~\cite{oppenheimer2006control}. As an example, in a turboprop with tractor propellers ahead of the wing, the propeller slipstream raises the local dynamic pressure over ailerons and flaps, therefore the effectiveness of aerodynamic surfaces changes with propeller activity as demonstrated in~\cite{soikkeli2023cascaded}. A time frozen linear control effectiveness matrix can not represent these dependencies, so purely linear allocation risks mismatched effectiveness, degraded optimality, and even infeasibility as conditions shift.

Instead of treating the control effectiveness matrix as fixed, it can be updated 
at each control cycle using an onboard model of the system~\cite{durham2017aircraft}. 
This approach is typical for incremental control allocation: at every cycle the 
onboard model is sampled at small perturbations around the current operating point, 
finite difference slopes are computed, and these partial derivatives are assembled 
into an updated control effectiveness matrix. Since each column of control effectiveness matrix
represents the partial derivative of the generalized forces and moments with respect 
to a single effector, the onboard model must be perturbed and evaluated separately 
for each of the $m$ effectors at every control update. This per-column finite 
difference procedure scales directly with the number of effectors and introduces 
a significant computational burden, particularly in highly overactuated platforms 
where $m$ is large. Furthermore, the onboard model often relies on vehicle 
state dependent aerodynamic tables and computationally heavy subcomponent models are difficult to interpret, and are highly 
sensitive to parameter variations in the numerical differentiation, potentially 
compromising transparency and robustness in the allocation loop.

Data driven control allocation methods, especially neural networks, are increasingly used to replace or augment classical linear allocators. Kang et al.\ train a neural allocator for a tailless aircraft to approximate a dynamic allocation scheme, preserving near optimal performance while avoiding heavy online optimization~\cite{kang2022development}. Similarly, Madruga et al.\ train a network offline to substitute the fixed allocation matrix on a small quadrotor, compensating aerodynamic effects without explicit inflow models~\cite{madruga2021aerodynamic}. In a complementary direction, Chen et al.\ integrate an adaptive neural attitude controller with a constrained allocation layer, providing robustness to uncertainties and time varying disturbances~\cite{chen2015constrained}. Building on these data driven allocators, reinforcement learning (RL) methods learn control allocation policies directly from interaction. For example, on the highly over actuated aircraft, RL has been used to distribute required input across 13 effectors for longitudinal control without a preidentified effectiveness matrix, while simultaneously reducing effector activity~\cite{de2019reinforcement}. Although neural network and reinforcement learning allocators can perform well, they are fundamentally function approximators that behave black box. The mapping from state to command is not transparent, which complicates verification, certification, and safety case construction. Behavior under distribution shift or rare corner cases are difficult to predict or bound, secondary objectives and constraints are enforced indirectly through loss function shaping, and failure diagnosis or accountability after incidents is hindered by limited interpretability. In safety critical settings, this explainability often outweighs the gains in empirical performance.

\begin{figure*}[ht]
    \centering
    \includegraphics[width=0.95\textwidth]{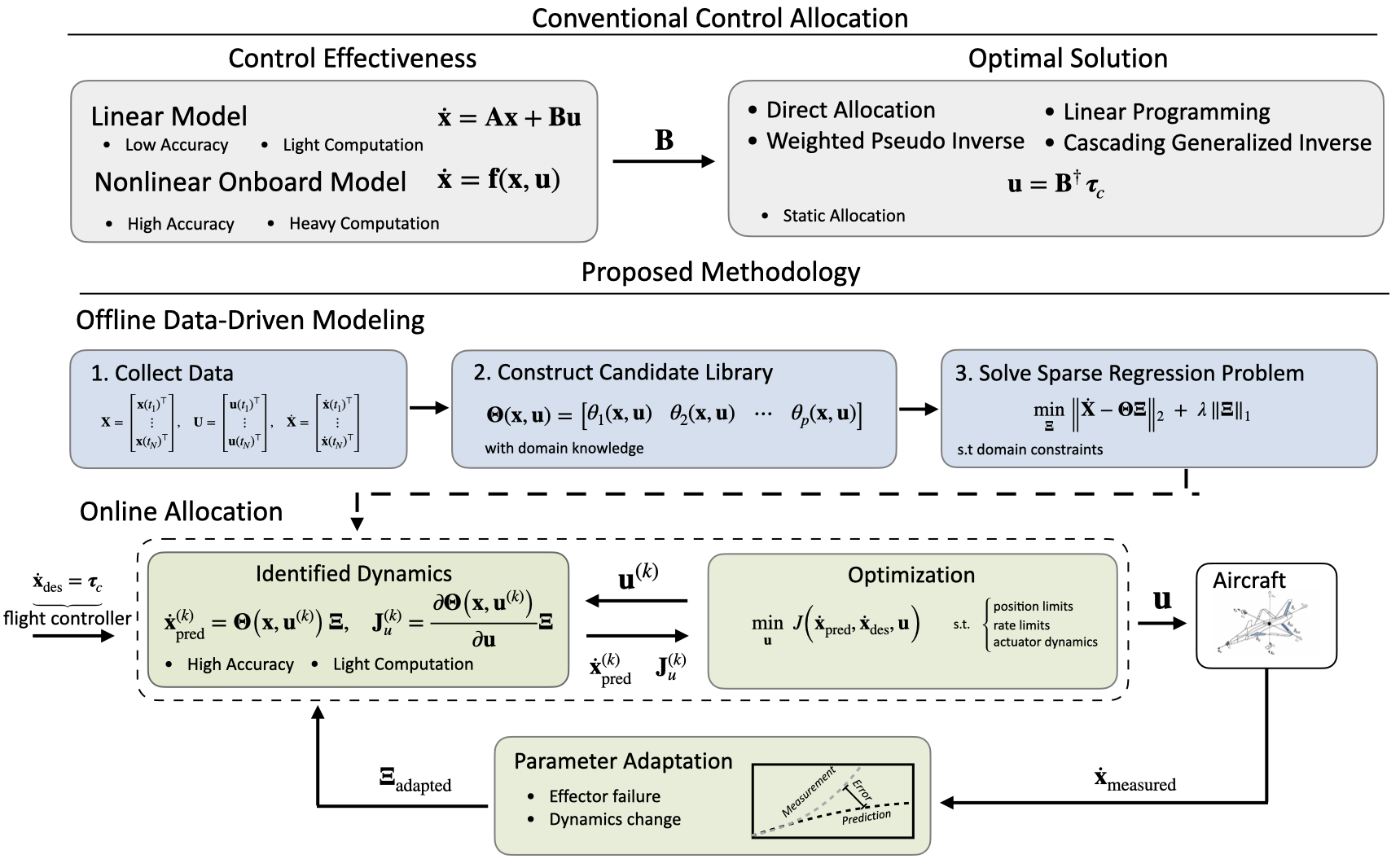}
    \caption{Conceptual diagram of the proposed control allocation framework. 
    The conventional linear onboard model relies on a constant effectiveness matrix 
    that loses validity under strong nonlinear couplings, while the high fidelity 
    nonlinear onboard model incurs a computational burden that limits real time use. 
    Both are replaced by a sparse analytical model identified from flight data, which 
    enables fast nonlinear optimization for control allocation with online parameter 
    adaptation.}\label{fig:framework}
\end{figure*}

Motivated by the drawbacks of black box approaches, we turn to a transparent, 
physics revealing alternative. Brunton et al.~\cite{brunton2016discovering} 
introduced the idea that governing equations can be recovered directly from 
data by expressing the unknown dynamics as a sparse linear combination drawn 
from a large library of candidate nonlinear functions of the states and inputs, 
a framework they termed Sparse Identification of Nonlinear Dynamics (SINDy). 
The result is a compact, interpretable set of governing equations whose 
structure can incorporate physics based constraints. A detailed mathematical formulation is provided in Section~\ref{sec:identification}. In recent years, SINDy has attracted substantial attention across disciplines, with applications ranging from discovering equations in fluid flows~\cite{fukami2021sparse} to identifying nonlinear aircraft dynamics~\cite{liu2025data}, modeling multirotor drones~\cite{manaa2024data}~\cite{chu2020discovering}, and capturing coupled disturbance and actuation dynamics on a rolling delta wing aircraft~\cite{deltawingsindy}. This breadth of successful applications suggests that SINDy is a natural candidate for learning the nonlinear control effectiveness mapping directly from limited data.

Taken together, the literature highlights a persistent tradeoff: high fidelity nonlinear models and advanced nonlinear allocators improve tracking under strong couplings, but they are often too computationally demanding for real time use, while linearized or fixed effectiveness formulations are efficient but lose accuracy as the effectiveness varies across the envelope. Data driven models can improve prediction, yet their limited transparency complicates validation and offers little insight into the underlying physics. These drawbacks motivate a middle ground in which data are used to learn an explicit, interpretable model that can be evaluated and differentiated efficiently and can be updated when the plant changes. Accordingly, this work pursues three tightly coupled objectives:

\begin{itemize}
    \item \textbf{Modeling fidelity}: the control effectiveness mapping must capture the nonlinear, state dependent  aerodynamic effects and effector interactions that arise during aggressive  maneuvering, so that the commanded virtual control is tracked accurately across a wide range of flight conditions.
    
    \item \textbf{Real time compatibility}: the mapping must be compact enough to 
    be evaluated and differentiated efficiently at each control update, without 
    requiring finite difference perturbations or repeated evaluation of a full 
    onboard aerodynamic model, so that the allocation solver remains feasible within 
    the available computational budget.

    \item \textbf{Model robustness}: the control effectiveness mapping must adapt 
    gracefully when the plant deviates from the conditions under which it was 
    identified, including mass and inertia variations caused by fuel consumption or 
    refueling, configuration changes, and partial or total loss of an effector, 
    so that allocation accuracy and tracking performance are preserved without 
    requiring a full reidentification.
\end{itemize}

To satisfy the combined objectives of robustness, transferability, and real time adaptability, this paper offers three specific contributions:

\begin{itemize}
    \item Using a constrained Sparse Identification of Nonlinear Dynamics (SINDy) approach, the control effectiveness mapping is learned offline from representative flight data, yielding a transparent, physics consistent model of control effectiveness mapping without relying on computationally heavy onboard models, linear approximations, or black box neural networks.
    \item These identified equations are subsequently embedded in nonlinear 
    optimization frameworks, which exploits their accuracy to generate actuator 
    commands that precisely realize the virtual control demands received from 
    the flight controller.
    \item  The same identification routine also runs online, continuously monitoring any mismatch between predicted and measured dynamics and refreshing its parameters whenever deviations emerge, thereby preserving model accuracy throughout changing flight conditions.
\end{itemize}

Closest work in the literature is the spline based nonlinear dynamic inversion controller of Tol et al.~\cite{tol2014nonlinear}, which replaces polynomial onboard models with multivariate simplex splines defined in local barycentric coordinates. Both approaches embed an explicit, differentiable, data driven model into the allocation loop to capture nonlinear control effectiveness. The key differences are that the spline representation is a general purpose piecewise approximator without direct physical meaning, constructed from a large dataset from known wind tunnel tables, whereas our identified model recovers a sparse, interpretable governing structure from time series trajectory data that can be obtained directly from flight experiments, making it more readily transferable across platforms and applications. Additionally, our framework supports online coefficient adaptation and extends the allocation to dynamic allocation formulation with explicit actuator dynamics propagation, neither of which is addressed in the spline formulation.

The remainder of this paper is organized as follows. Section~\ref{sec:problem_statement} formulates the control allocation problem, defines the actuator constraints, and introduces the baseline onboard and linear allocation models used for comparison. Section~\ref{sec:identification} presents the constrained SINDy identification framework, including the physics informed library construction and the offline training procedure. Section~\ref{sec:control_allocation} describes how the identified dynamics are embedded into nonlinear control allocation and introduces the adaptation mechanism for maintaining accuracy under changing conditions and failures. Section~\ref{sec:results} reports simulation results on agile maneuvers, including Monte Carlo validation, computational timing, and failure case evaluations. Finally, Section~\ref{sec:conclusion} concludes the paper and outlines directions for future work.

\section{Problem Statement}\label{sec:problem_statement}

In this section, we formulate the nonlinear control allocation problem for an overactuated system and describe the proposed workflow, which consists of learning an explicit control effectiveness mapping from offline data and using this mapping online to compute feasible actuator commands that realize the requested virtual control from the flight controller. 

The overall methodology is illustrated in Fig.~\ref{fig:framework}. In the offline phase, representative maneuver data are collected from a high fidelity nonlinear aircraft model spanning a range of flight conditions, actuator deflections, and excitation levels to ensure sufficient richness in the state, input, and rate channels. Using these data, a structured candidate model set is constructed from measured states $x$, control inputs $u$, state variables (e.g., $V$, $\alpha$, $\beta$), and selected nonlinear interaction terms. A sparse regression procedure is then used to identify a parsimonious continuous time input to state model that is consistent with the observed dynamics. In the online phase, the learned explicit dynamics are used as the prediction mapping inside a nonlinear allocation problem. At each control update, the allocator receives the current state and scheduling variables and evaluates the model to predict the body rate response for a candidate actuator vector. The optimizer then iteratively adjusts the actuator commands to reduce the mismatch between the predicted response and the commanded body rate demand, while simultaneously penalizing actuator usage and variations from a reference or previous command to avoid unnecessary control activity. Hard bounds and rate limits are enforced directly through simple projections or constraint handling within the solver, ensuring that the returned command is feasible by construction. A key advantage of using an explicit analytical model is that its sensitivities with respect to actuator inputs are available in analytical form. These analytical Jacobians are supplied to the nonlinear solver to compute reliable search directions, particularly in strong nonlinear regions.

To validate the eﬀectiveness of our methodology, we use the ADMIRE (Aero-Data Model in a Research Environment) simulation~\cite{Admire_report}, a standard benchmark in the aircraft control allocation literature and the primary platform in~\cite{durham2017aircraft}. ADMIRE is a publicly available, nonlinear MATLAB/Simulink model of a generic small, single seat, single engine delta canard fighter aircraft, providing a high fidelity testbed with multiple redundant aerodynamic effectors and pronounced nonlinear flight regime behavior. The ADMIRE model comprises a 12-state, six-degree-of-freedom airframe coupled with aerodynamic data tables, an engine model, and first order actuator and sensor dynamics, including deflection and rate limits as well as a representative computational delay. The operating envelope extends to Mach 1.2 and 6~km altitude, with extended aerodynamic data enabling simulation in high angle of attack regimes. The aircraft is overactuated with seven independent aerodynamic effectors: right canard deflection $(\delta_{rc})$, left canard deflection $(\delta_{lc})$, right outboard elevon deflection $(\delta_{roe})$, right inboard elevon deflection $(\delta_{rie})$, left inboard elevon deflection $(\delta_{lie})$, left outboard elevon deflection $(\delta_{loe})$, and rudder deflection $(\delta_{r})$. This control redundancy is exploited by the allocator to satisfy actuator position and rate limits, accommodate secondary objectives, and retain control authority under effector degradations or failures.

\begin{figure}[t]
    \centering
    \includegraphics[width=0.75\linewidth]{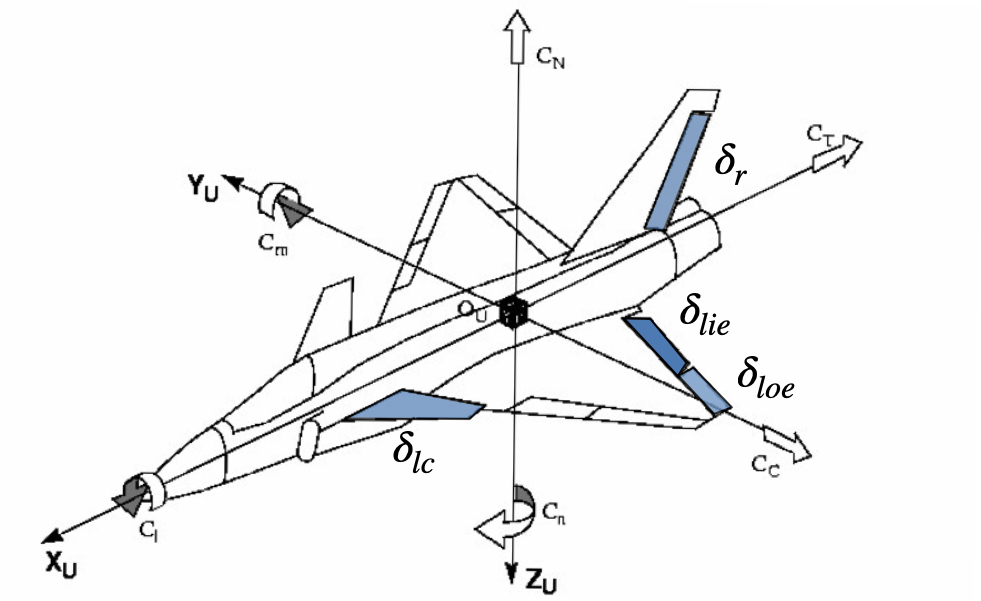}
    \caption{ADMIRE aircraft model. The aerodynamic control surfaces are highlighted in color to indicate the effectors used in this study, including their symmetric pairs.}
    \label{fig:admire3d}
\end{figure}

\begin{table}[t]
\centering
\caption{Configuration parameters of the ADMIRE aircraft.}
\label{tab:nominal_config}
\setlength{\tabcolsep}{10pt}
\renewcommand{\arraystretch}{1.3}
\begin{tabular}{l 
                l
                l}
\toprule
\textbf{Parameter} & \textbf{Value} & \textbf{Unit} \\
\midrule
Wing area,\ $S$               & 45.00  & $\mathrm{m}^{2}$          \\
Wing span,\ $b$               & 10.00  & $\mathrm{m}$              \\
Mean wing chord,\ $\bar{c}$   & 5.20  & $\mathrm{m}$              \\
Mass,\ $m$                    & 9100   & $\mathrm{kg}$             \\
\midrule
$I_{x}$                       & 21000  & $\mathrm{kg\,m^{2}}$      \\
$I_{y}$                       & 81000  & $\mathrm{kg\,m^{2}}$      \\
$I_{z}$                       & 101000 & $\mathrm{kg\,m^{2}}$      \\
$I_{xz}$                      & 2500  & $\mathrm{kg\,m^{2}}$      \\
\bottomrule
\end{tabular}
\end{table}

An illustration of the ADMIRE aircraft body frame and the selected control surfaces is shown in Fig.~\ref{fig:admire3d}, and the corresponding nominal configuration parameters used in this study are summarized in Table~\ref{tab:nominal_config}.

\section{Control Effectiveness Identification}\label{sec:identification}
Reliable nonlinear control allocation requires an explicit model that captures how the controlled variables respond to actuator inputs across changing operating conditions. While high fidelity onboard models can provide this mapping, they are often too costly for real time optimization, and local linearization may not remain valid in strongly nonlinear regimes. This motivates a data driven identification step that learns a compact, interpretable set of equations directly from representative flight data and can be evaluated efficiently during online allocation. In this work, we adopt a sparse regression approach to infer continuous time dynamics from measured trajectories, yielding an explicit model structure that supports systematic incorporation of domain knowledge.

Sparse Identification of Nonlinear Dynamics (SINDy) is a system identification framework that constructs an explicit continuous time model directly from time resolved trajectories~\cite{brunton2016discovering}. Given measurements of the state $\mathbf{x}(t)$ and input $\mathbf{u}(t)$,  SINDy assumes that the state derivatives can be written as a sparse linear combination of candidate nonlinear features,
and seeks a compact coefficient matrix $\mathbf{\Xi}$ whose nonzero entries identify the few terms that govern the observed evolution. To this end, SINDy evaluates a user defined library $\mathbf{\Theta}(\cdot)$ built from nonlinear basis functions of the measured signals (e.g., polynomial interactions, trigonometric terms, and other selected functions), and applies sparse regression to fit these features to the measured time derivatives $\dot{\mathbf{x}}(t)$ (or calculated derivatives $\dot{\mathbf{x}}(t_k)$ via numerical differentiation), retaining only the active terms required for accurate prediction. This produces a parsimonious, interpretable model that is straightforward to evaluate and differentiate, making it well suited for control allocation problem. We begin from the standard continuous time state space form,
We consider a continuous time nonlinear system driven by measured control inputs,
\begin{equation}
\dot{\mathbf{x}}(t)=\mathbf{f}\!\bigl(\mathbf{x}(t),\mathbf{u}(t)\bigr),
\qquad
\mathbf{x}\in\mathbb{R}^{n},\;
\mathbf{u}\in\mathbb{R}^{m},
\end{equation}
where $\mathbf{x}(t)$ is the $n$-dimensional state vector and $\mathbf{u}(t)$ is the $m$-dimensional actuator command. Given a set of time samples $\{ t_k \}_{k=1}^{N}$, we collect the measured trajectories into data matrices
\begin{align}
\mathbf{X} &=
\begin{bmatrix}
\mathbf{x}(t_1)^\top\\[-2pt]
\vdots\\[-2pt]
\mathbf{x}(t_N)^\top
\end{bmatrix}\in\mathbb{R}^{N\times n}, \\
\mathbf{U} &=
\begin{bmatrix}
\mathbf{u}(t_1)^\top\\[-2pt]
\vdots\\[-2pt]
\mathbf{u}(t_N)^\top
\end{bmatrix}\in\mathbb{R}^{N\times m}, \\
\dot{\mathbf{X}} &=
\begin{bmatrix}
\dot{\mathbf{x}}(t_1)^\top\\[-2pt]
\vdots\\[-2pt]
\dot{\mathbf{x}}(t_N)^\top
\end{bmatrix}\in\mathbb{R}^{N\times n}.
\end{align}

where $\dot{\mathbf{x}}(t)$ is obtained from measurements or numerical differentiation of $\mathbf{x}(t)$.

SINDy represents the unknown nonlinear mapping using a library of candidate basis functions evaluated on the measured data. Specifically, define a row vector of $p$ candidate functions
\[
\boldsymbol{\theta}\bigl(\mathbf{x},\mathbf{u}\bigr)
=
\begin{bmatrix}
\theta_{1}(\mathbf{x},\mathbf{u}) &
\theta_{2}(\mathbf{x},\mathbf{u}) &
\cdots &
\theta_{p}(\mathbf{x},\mathbf{u})
\end{bmatrix}
\in\mathbb{R}^{1\times p},
\]
where each $\theta_i(\cdot)$ is a chosen nonlinear feature. Evaluating this row at every time sample and stacking the results yields the library matrix
\[
\mathbf{\Theta}(\mathbf{X},\mathbf{U})
=
\begin{bmatrix}
\boldsymbol{\theta}(\mathbf{x}(t_1),\mathbf{u}(t_1))\\[-2pt]
\vdots\\[-2pt]
\boldsymbol{\theta}(\mathbf{x}(t_N),\mathbf{u}(t_N))
\end{bmatrix}
\in\mathbb{R}^{N\times p}.
\]
The dynamics are then approximated as a linear combination of these candidate functions,
\[
\dot{\mathbf{X}} \approx \mathbf{\Theta}(\mathbf{X},\mathbf{U})\,\mathbf{\Xi},
\]
where $\mathbf{\Xi}\in\mathbb{R}^{p\times n}$ is the coefficient matrix. Each column of $\mathbf{\Xi}$ contains the coefficients for one state derivative, so sparsity in $\mathbf{\Xi}$ directly corresponds to selecting only a small subset of active library terms for each component of $\dot{\mathbf{x}}$.

Given the data matrices $\dot{\mathbf{X}}$ and $\boldsymbol{\Theta}$, identifying the governing equations reduces to estimating the coefficient matrix $\mathbf{\Xi}$ such that the library output $\boldsymbol{\Theta}\mathbf{\Xi}$ matches the measured derivatives while retaining only a small number of active terms. This is achieved by solving a regularized regression problem that balances prediction accuracy against model complexity through a sparsity promoting penalty:
\begin{equation}
\min_{\mathbf{\Xi}\in\mathbb{R}^{p\times n}}
\;\Bigl\lVert\dot{\mathbf{X}} - \boldsymbol{\Theta}\mathbf{\Xi}\Bigr\rVert_{2}
\;+\;
\lambda\,\|\mathbf{\Xi}\|_{1},
\end{equation}
where $\dot{\mathbf{X}}\in\mathbb{R}^{N\times n}$ contains the sampled state derivatives $\dot{\mathbf{x}}(t_k)$, $\boldsymbol{\Theta}\in\mathbb{R}^{N\times p}$ is the evaluated library matrix, and $\lambda\ge 0$ controls the tradeoff between data-fit and sparsity by penalizing the $L_1$ norm of $\mathbf{\Xi}$.

A variety of sparse regression strategies have been proposed in the literature to solve this identification problem, including $L_1$-based methods and thresholded least squares variants. While the $L_1$-regularized formulation above provides a direct sparsity mechanism, in practice we found that standard solvers such as LASSO (Least Absolute Shrinkage and Selection Operator) or STLSQ (Sequentially Thresholded Least Squares) are sensitive to the sparsity penalty: with small penalties they retain many small but nonzero coefficients, and with larger penalties they often discard dynamically important terms and degrade fit. For the high fidelity nonlinear aircraft model considered here, this tradeoff produced models that were either insufficiently sparse or insufficiently accurate, motivating a more robust identification procedure.

To address this, we adopt sparse relaxed regularized regression (SR3)~\cite{SR3}, which improves robustness by decoupling the data fitting coefficients from the sparsity enforced coefficients through an auxiliary variable. In addition, SR3 admits explicit constraints on the coefficient matrix, allowing domain knowledge to be enforced directly during regression (e.g., removing nonphysical cross terms or enforcing expected damping signs), rather than relying on sparsity alone to recover the correct structure. This results in models that are simultaneously parsimonious, better conditioned, and more consistent with known physics, which is critical when the identified dynamics are later embedded into nonlinear solvers.

Specifically, SR3 solves
\begin{align}
\min_{\Xi,\,W} \;
&\frac{1}{2}\big\|\dot{X} - \Theta\,\Xi \big\|_F^2
\;+\; \frac{\kappa}{2}\big\|\Xi - W\big\|_F^2
\;+\; \lambda \,\|W\|_{1},
\end{align}
where $W \in \mathbb{R}^{p \times n}$ is a relaxed sparsity-promoting copy of $\Xi$, $\kappa > 0$ sets the strength of the relaxation, and $\lambda > 0$ controls sparsity~\cite{SR3}.

All identification experiments in this work are implemented using the open source \texttt{PySINDy} toolbox~\cite{desilva2020}, which provides a unified interface for constructing candidate libraries, estimating time derivatives, and solving sparse regression problems with multiple optimizers, including SR3 and constrained variants. We use \texttt{PySINDy} to (i) assemble the physics informed library, (ii) fit the coefficient matrix using the selected sparse regression solver and constraints, and (iii) export the identified continuous time model for integration into the control allocation routines.

\subsection{Incorporating Domain Knowledge into the Library}
When assembling the candidate function library for SINDy, domain knowledge is indispensable. A naive polynomial expansion grows rapidly with state and input dimension, leading to large, poorly conditioned dictionaries that admit many spurious correlations and often yield non robust identified models. In our setting, the identified model is intended for control allocation, where the central object is the control effectiveness mapping that relates actuator deflections to generated forces and moments, and equivalently to the resulting angular accelerations. After extensive trials with generic polynomial candidate terms that failed to recover consistent and transferable models, we found that embedding the known structure of aircraft rotational dynamics into the library significantly improves recovery of the governing equations. A similar observation is reported in~\cite{liu2025eisindyc}, where integrating prior flight mechanics knowledge into the candidate library improves  identification accuracy and generalization under limited data conditions. In particular, the compact formulation of Stevens and Lewis expresses the angular accelerations as an explicit combination of rigid body rate coupling terms and the total aerodynamic moments~\cite{stevens2015aircraft}:

\begin{align}
\dot{p} &= (C_{1}r + C_{2}p)\,q + C_{3}L + C_{4}N  \notag \\ 
\dot{q} &= C_{5}pr - C_{6}(p^{2}-r^{2}) + C_{7}M \\
\dot{r} &= (C_{8}p - C_{2}r)\,q + C_{4}L + C_{9}N .  \notag
\end{align}

Here, $L$, $M$, and $N$ denote the total body axis moments about the roll, pitch, and yaw axes, respectively, and include both aerodynamic and propulsive contributions. The coefficients $C_i$ depend only on the inertia terms and can be computed from $(I_{xx},I_{yy},I_{zz},I_{xz})$ via

\begin{align}
\Gamma &= I_{xx}I_{zz}-I_{xz}^{2}  \notag \\
C_{1} &= \frac{(I_{yy}-I_{zz})I_{zz}-I_{xz}^{2}}{\Gamma} \notag
&\qquad
C_{4} &= \frac{I_{xz}}{\Gamma} \notag \\
C_{2} &= \frac{(I_{xx}-I_{yy}+I_{zz})I_{xz}}{\Gamma} \notag
&\qquad
C_{5} &= \frac{I_{zz}-I_{xx}}{I_{yy}} \notag \\
C_{3} &= \frac{I_{zz}}{\Gamma}
&\qquad
C_{6} &= \frac{I_{xz}}{I_{yy}} \\ \notag
C_{8} &= \frac{I_{xx}(I_{xx}-I_{yy})+I_{xz}^{2}}{\Gamma} \notag
&\qquad
C_{7} &= \frac{1}{I_{yy}} \\ \notag
C_{9} &= \frac{I_{xx}}{\Gamma} \notag
\end{align}

These inertia based coefficients are used to normalize the candidate library terms (e.g., moment contributions to $\dot{p},\dot{q},\dot{r}$), such that the resulting features have comparable magnitudes. This normalization improves numerical conditioning of the sparse regression and optimization, making the identification problem easier to solve and reducing sensitivity to ill-scaled terms. The decomposition provides a principled template for the SINDy library: the quadratic coupling products $(pq,\,pr,\,qr,\,p^2,\,r^2)$ are known to appear from rigid body dynamics, while the moments $(L,M,N)$ aggregate aerodynamic and propulsive contributions. Consequently, instead of allowing arbitrary nonlinear combinations of $(p,q,r)$ and control deflections, we explicitly include the physically motivated candidate functions. Specifically, aerodynamic moments are modeled using dynamic pressure scaling, reference geometry, and control deflections:

\begin{align}
L &\approx \bar{q} S b \; C_{\ell}(\alpha,\beta,\boldsymbol{\delta}), \notag \\ 
M &\approx \bar{q} S c \; C_{m}(\alpha,\beta,\boldsymbol{\delta}), \\
N &\approx \bar{q} S b \; C_{n}(\alpha,\beta,\boldsymbol{\delta}), \notag \\
\bar{q} &= \tfrac{1}{2}\rho V^2 . \notag
\end{align}

where $\boldsymbol{\delta}$ collects the effector deflections. This motivates a cascaded, domain informed dictionary that separates known 
rigid body structure from unknown aerodynamic mappings. The candidate library 
is constructed as

\begin{equation}
\Theta(\mathbf{x}, \boldsymbol{\delta}) = 
\bigl[\,\theta_1(\mathbf{x},\boldsymbol{\delta}),\;
\theta_2(\mathbf{x},\boldsymbol{\delta}),\;
\dots,\;
\theta_p(\mathbf{x},\boldsymbol{\delta})\,\bigr] \in \mathbb{R}^{1 \times p},
\end{equation}

where each $\theta_i$ is a physically motivated candidate function drawn from 
the following groups:

\smallskip
\noindent
\begin{tabular}{@{}ll@{}}
Body rates          & $p^{k},\; q^{k},\; r^{k}\quad (k=1,\dots,K)$ \\[3pt]
Rigid-body products & $C_2\,pq,\; C_1\,pr,\; C_2\,qr,\;\dotsc$ \\[3pt]
Rate damping        & $\tfrac{b}{2V}\,p,\;\; \tfrac{\bar{c}}{2V}\,q,\;\dotsc$ \\[5pt]
Control effectiveness & $\tfrac{\bar{q}Sb}{I_x}\,\delta_1,\;\;
                         \tfrac{\bar{q}S\bar{c}}{I_y}\,\delta_2,\;\;
                         \tfrac{\bar{q}Sb}{I_z}\,\delta_3,\;\dotsc$ \\[5pt]
Control couplings   & $\delta_1\delta_2,\; \delta_1\delta_3,\;
                       \delta_2\delta_3,\;\dotsc$ \\[3pt]
Aerodynamic angles  & $\alpha,\;\beta,\;\sin\alpha,\;\cos\beta,\;\dotsc$ \\[3pt]
Dynamic pressure    & $\bar{q}=\tfrac{1}{2}\rho V^{2}$
\end{tabular}
\smallskip


In this way, SINDy is guided toward interpretable models that respect established 
physics: inertia driven coupling is embedded a priori through $(C_1,\dots,C_9)$ 
and the rigid body product terms in $\Theta$, while the control effectiveness and 
aerodynamic angle terms primarily capture the state dependent aerodynamic 
contributions contained in $C_{\ell}, C_{m}, C_{n}$. This structure reduces the 
search space, improves identifiability, and yields a coefficient matrix $\Xi$ 
whose nonzero entries have direct physical meaning.

\subsection{Data Collection}
We instrumented the agile aircraft to execute a curated set of high excitation maneuvers designed to excite nonlinear effects. Simulations were run with $dt=0.01\,\mathrm{s}$ and synchronized logging of the state vector
\begin{equation}
\mathbf{x}(t)=\begin{bmatrix} p(t) & q(t) & r(t) \end{bmatrix}^{\top},
\end{equation}
together with the full input vector
\begin{equation}
\mathbf{u}(t)=\begin{bmatrix} \delta_{rc} & \delta_{lc} & \delta_{roe} & 
\delta_{rie} & \delta_{lie} & \delta_{loe} & \delta_{r} \end{bmatrix}^{\top},
\end{equation}
where both the commanded deflections $\mathbf{u}_{\mathrm{cmd}}(t)$ and the 
realized actuator deflections $\mathbf{u}_{\mathrm{act}}(t)$ are recorded to 
capture the effect of actuator dynamics, and variables 
$(V(t),\, \alpha(t),\, \beta(t))$ representing airspeed, angle of attack, and 
sideslip angle respectively. The maneuver set was repeated across multiple flight 
conditions spanning several altitudes and Mach numbers to capture variations in 
dynamic pressure and control effectiveness. In total, $200$ trajectories of 
$10$\,s duration were generated. Half of the dataset was used for system 
identification (training), and the remaining half was held out for evaluation, 
both in open loop prediction tests and in closed loop control tasks to assess 
performance.

\begin{figure}[ht]
    \centering
    \includegraphics[width=0.45\textwidth]{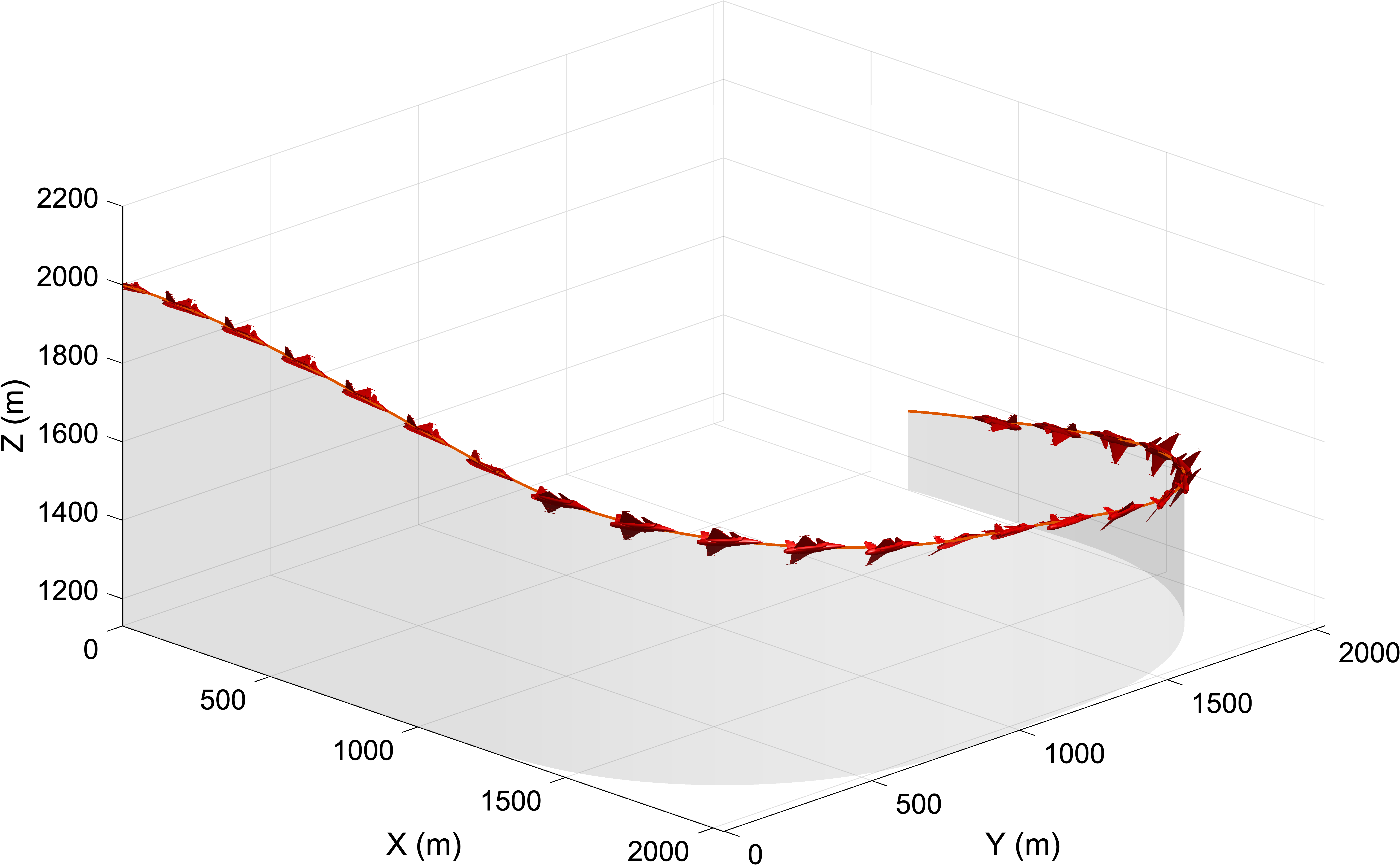}
    \caption{A representative trajectory from the dataset}\label{fig:example_trajectories}
\end{figure}
Additionally, the maneuvers are designed to both excite individual control surface deflections and combined surface deflections, enabling identification of each surface’s individual contribution as well as nonlinear effects. Example trajectory from these maneuvers are shown in Fig~\ref{fig:example_trajectories}.

\subsection{Identification Results}

This section reports the system identification results for the aircraft’s rotational dynamics using the test dataset. The identified model is learned offline from the training trajectories using constrained SR3, with physics informed equality and inequality constraints and an $L_1$ thresholding rule. To improve robustness of term selection and reduce sensitivity to individual maneuvers, we employ an ensemble strategy with 50 independently identified models obtained. The final coefficient matrix is obtained by aggregating the ensemble, using the mean coefficient values. In our experiments, disabling library ensembling led to unstable sparsity patterns and inconsistent coefficients across maneuvers, which translated to noticeably poorer generalization on the test set.

To evaluate whether the learned dynamics faithfully reproduce the plant behavior, we perform an open loop validation over the 100 test trajectories with randomized initial conditions and distinct control input sequences. For each trial, the identified continuous time model is initialized at the measured initial state and integrated forward using the recorded test inputs with a fourth order Runge Kutta method, producing predicted body rate trajectories. The predicted roll, pitch, and yaw rates are then compared directly against the corresponding test trajectories to quantify generalization across operating conditions, including variations in altitude, airspeed, and aerodynamic angles.

\begin{figure}[ht]
    \centering
    \includegraphics[width=0.45\textwidth]{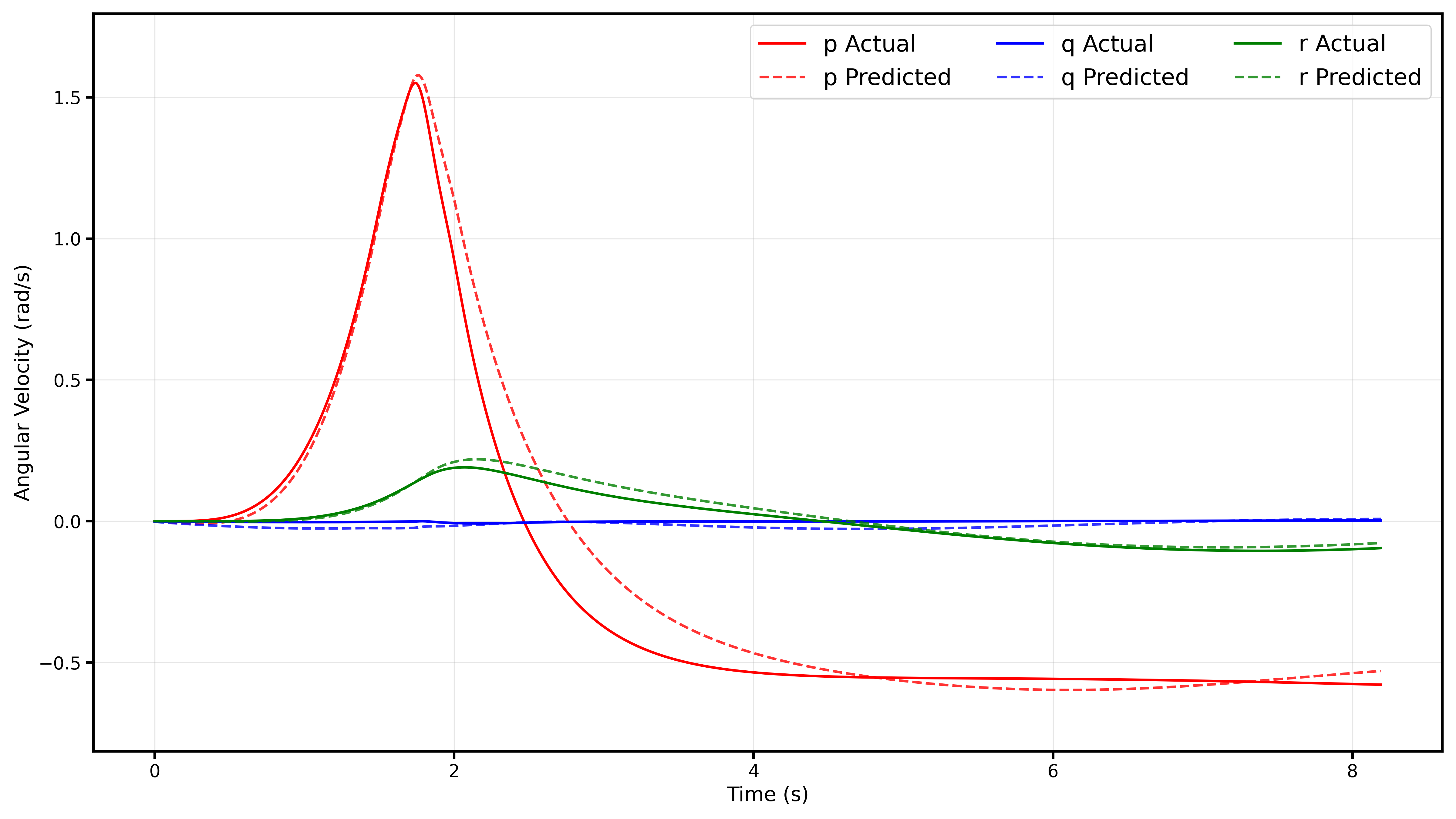} 
    \includegraphics[width=0.45\textwidth]{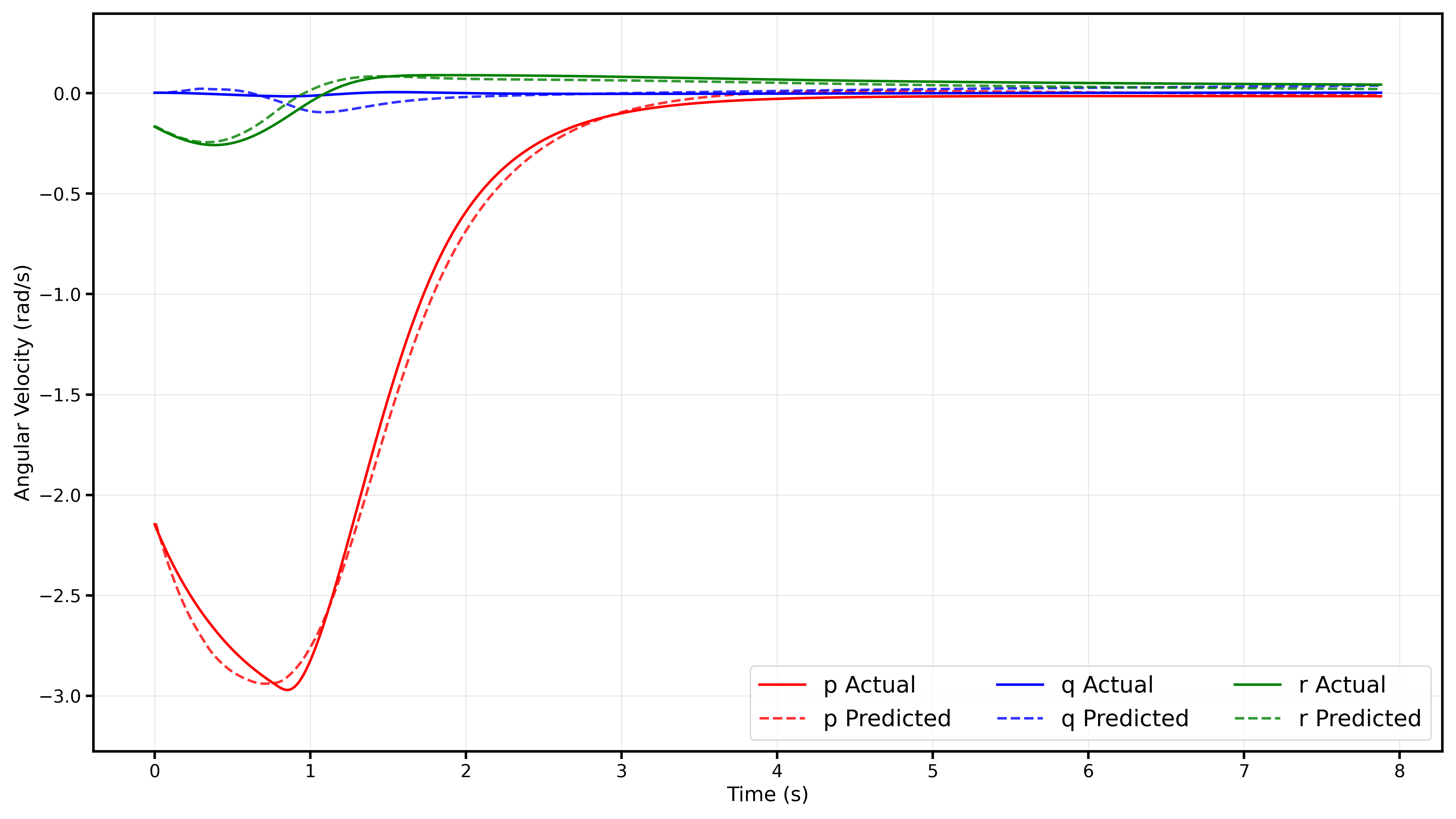}
    \caption{Body rate prediction on test trajectories. Solid lines show the measured/test data for roll, pitch, and yaw rates, while dashed lines show the rates predicted by the SINDy identified dynamics integrated forward from the same initial condition.}\label{fig:2x2}
\end{figure}

Figure~\ref{fig:2x2} shows representative examples from the Monte Carlo test set. Even under highly nonlinear, high excitation agile maneuvers, the learned dynamics reproduce the measured body rate trajectories with close agreement in open loop. This is particularly relevant for onboard use: despite its compact form and low computational cost, the identified model captures the dominant nonlinear effects of the true plant dynamics, making it suitable as a real time prediction and allocation model.

\begin{table*}[htbp]
\centering
\caption{Open loop model prediction error over 100 Monte Carlo test maneuvers 
         for different sparsity levels.}\label{tab:sparsity_error_pqr}
\setlength{\tabcolsep}{10pt}
\renewcommand{\arraystretch}{1.3}
\begin{tabular}{l 
                S[table-format=2.0] 
                S[table-format=3.2] 
                S[table-format=1.4] 
                S[table-format=1.4] 
                S[table-format=1.4]
                S[table-format=1.4]}
\toprule
\textbf{Sparsity Level} 
    & \textbf{\# Terms} 
    & \textbf{Comp.\ Time (s)} 
    & \textbf{$p$-RMSE}
    & \textbf{$q$-RMSE}
    & \textbf{$r$-RMSE}
    & $\overline{\textbf{RMSE}}$ \\
\midrule
High   $(\lambda = 0.1)$       & 20 &  35.58 & 0.0560 & 0.0148 & 0.0026 & 0.0077 \\
Medium $(\lambda = 10^{-3})$   & 38 &  75.27 & 0.0017 & 0.0293 & 0.0309 & 0.0009 \\
Low    $(\lambda = 10^{-5})$   & 42 & 140.17 & 0.0096 & 0.0149 & 0.0072 & 0.1060 \\
\bottomrule
\end{tabular}
\end{table*}

Table~\ref{tab:sparsity_error_pqr} quantifies the open loop prediction accuracy over the $100$ Monte Carlo test maneuvers for low, medium, and high sparsity settings, reporting the RMSE for each axes $(p,q,r)$ together with the mean RMSE across axes, the number of active terms, and the corresponding evaluation time. As expected, reducing sparsity (i.e., allowing more active terms) initially improves predictive accuracy by enabling the model to represent additional nonlinear terms present in the model. However, beyond a certain complexity, the identified models become less consistent across maneuvers and more sensitive to errors, which manifests as unstable coefficient estimates and reduced reliability despite the increased term count. Consequently, selecting the final model requires balancing accuracy against sparsity and numerical robustness. Based on this tradeoff, we proceed with the medium sparsity model for the remainder of the paper, as it provides sufficiently accurate open loop prediction while retaining a compact structure that is favorable for nonlinear allocation.

\begin{figure*}
    \centering
    \includegraphics[width=0.75\linewidth]{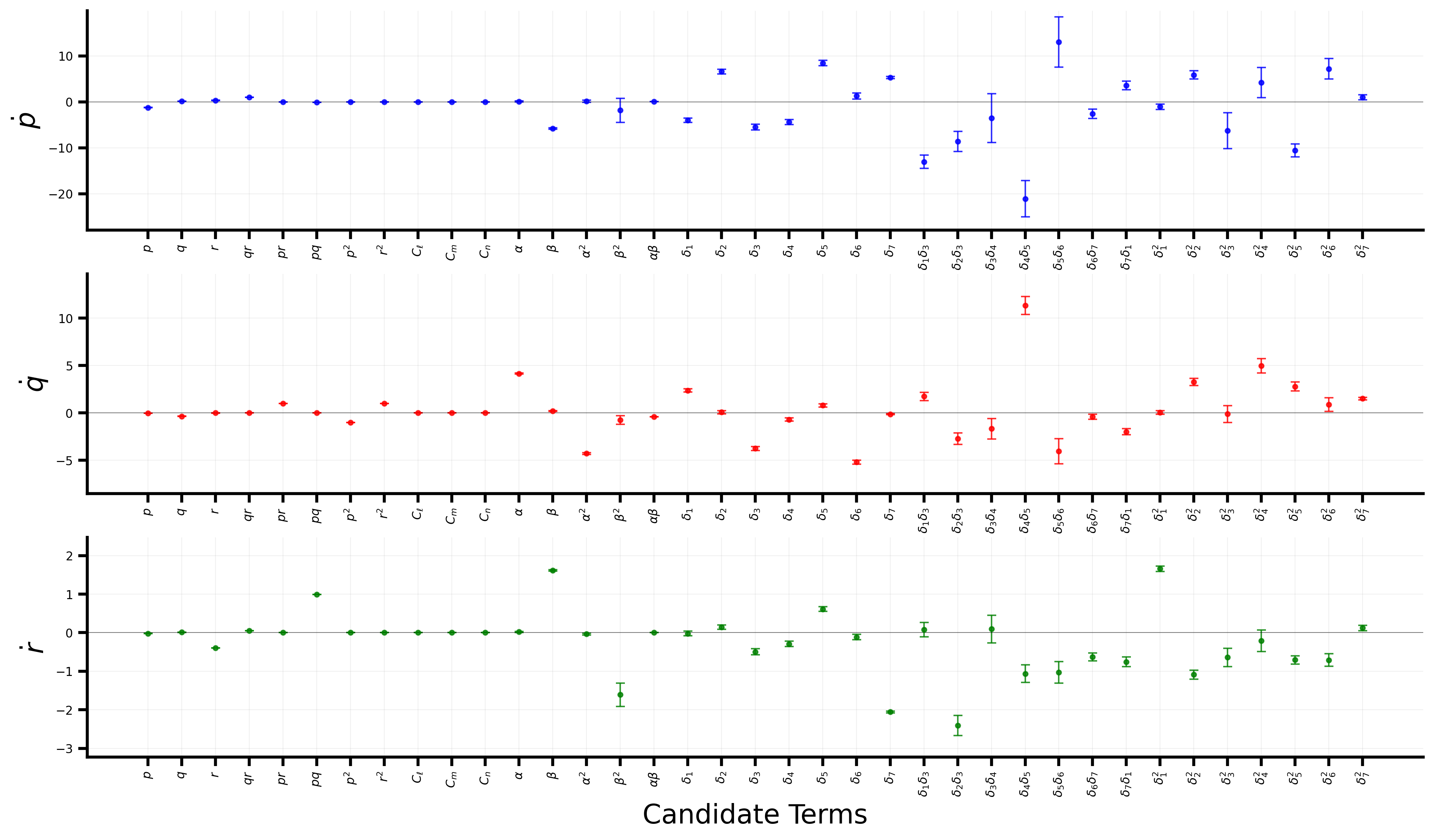}
\caption{Ensemble coefficients over 50 different identified models. For each library term, the dot denotes the ensemble mean coefficient, and the error bars indicate the corresponding standard deviation. The final model used in this work is formed by the ensemble mean coefficients.}\label{fig:probablistic_coeff}
\end{figure*}

\subsubsection{Interpretability and Physical Consistency of Identified Coefficients}

The proposed identification framework yields an explicit, sparse, and physically interpretable model. Beyond open loop trajectory predictions, interpretability is assessed by directly comparing the inertia driven coefficients implied by the identified dynamics with the known rigid body coefficients computed from the aircraft inertia tensor. Since the compact rotational equations parameterize angular accelerations through $(C_1,\dots,C_9)$, agreement between identified and physical values indicates that SINDy recovers not only an accurate predictor but also the correct governing structure. 

\begin{table}[htbp]
\centering
\caption{Comparison of known and identified coefficients.}\label{tab:rb_recovery_scaled}
\setlength{\tabcolsep}{8pt}
\renewcommand{\arraystretch}{1.3}
\begin{tabular}{l l 
                S[table-format=+1.4, table-comparator=true] 
                S[table-format=+1.4] 
                S[table-format=1.4]}
\toprule
\textbf{Eq.} 
    & \textbf{Term} 
    & \textbf{Known} 
    & \textbf{Identified} 
    & \textbf{Rel.\ Err.\ (\%)} \\
\midrule
\multirow{2}{*}{$\dot{p}$} 
    & $qr$ & -0.9582 & -0.9486 & 1.0000 \\
    & $pq$ &  0.0485 &  0.0492 & 1.4200 \\
\midrule
\multirow{2}{*}{$\dot{q}$} 
    & $pr$     &  0.9877 &  0.9778 & 1.0000 \\
    & $p^{2}$  & -0.0309 & -0.0306 & 0.0097 \\
    & $r^{2}$  &  0.0309 &  0.0306 & 0.0097 \\
\midrule
\multirow{2}{*}{$\dot{r}$} 
    & $pq$ & -0.5929 & -0.5870 & 0.0099 \\
    & $qr$ &  0.0485 & -0.0479 & 1.9870 \\
\bottomrule
\end{tabular}
\end{table}

Table~\ref{tab:rb_recovery_scaled} reports the known coefficients, the identified coefficients, and their relative errors, demonstrating how closely the learned model aligns with first principles rigid body dynamics.

\section{Nonlinear Control Allocation with Identified Control Effectiveness}\label{sec:control_allocation}
An explicit analytical model of the input to output dynamics provides substantial leverage for control allocation: the identified mapping can be locally linearized  to recover a state dependent effectiveness matrix, used in dynamic inversion to  compute actuator commands directly, or embedded in a nonlinear optimization  framework to account for constraints and secondary objectives. In the following, we present two complementary allocation strategies that exploit this analytical structure.

\subsection{Gradient Based Nonlinear Control Allocation}
Since the identified control effectiveness mapping $\mathbf{\Theta}(\mathbf{x},\mathbf{u})\,\mathbf{\Xi}$ is explicit and analytical, its Jacobian with respect to the actuator inputs is available in closed form and can be supplied directly to the iterative solver to accelerate convergence.

At each control update the flight computer provides a commanded angular acceleration $\dot{\mathbf{x}}_{\mathrm{des}}\in\mathbb{R}^{3}$, the desired rotational response the allocator must realize through the available effectors. For a candidate actuator command $\mathbf{u}\in\mathbb{R}^{m}$ the mapping predicts $\dot{\mathbf{x}}_{\mathrm{pred}}=\mathbf{\Theta}(\mathbf{x},\mathbf{u})\,\mathbf{\Xi}$, and allocation is posed as minimizing the weighted tracking mismatch with a regularization on actuator usage,
\begin{equation}
\min_{\mathbf{u}}\;
J(\mathbf{u})
=
\frac{1}{2}\bigl\|\dot{\mathbf{x}}_{\mathrm{pred}}-\dot{\mathbf{x}}_{\mathrm{des}}\bigr\|_{W}^{2}
+\frac{1}{2}\|\mathbf{u}\|_{R}^{2}.
\end{equation}

The problem is solved iteratively by projected gradient descent. The search direction follows from the gradient $\nabla_{\mathbf{u}}J = \mathbf{J}_u^\top W\,\mathbf{e} + R\,\mathbf{u}$, with tracking error $\mathbf{e}=\dot{\mathbf{x}}_{\mathrm{pred}}-\dot{\mathbf{x}}_{\mathrm{des}}$ and Jacobian $\mathbf{J}_u = (\partial\mathbf{\Theta}/\partial\mathbf{u})\,\mathbf{\Xi}$ evaluated analytically, so no finite differencing is required. Each iterate is projected onto the actuator bounds $[\mathbf{u}_{\min},\mathbf{u}_{\max}]$, and the step length is chosen by an Armijo backtracking line search, which guarantees sufficient decrease of $J$ at every step and removes the sensitivity to a fixed step size. The complete procedure is given in Algorithm~\ref{alg:pgd_ca}.

\begin{algorithm}[htbp]
\caption{Nonlinear Control Allocation via Gradient Descent with Armijo Line Search}\label{alg:pgd_ca}
\begin{algorithmic}[1]

\Require{Desired $\dot{\mathbf{x}}_{\mathrm{des},t} \in \mathbb{R}^n$}
\Require{Tracking weight $W \succeq 0$, effort weight $R \succ 0$}
\Require{Actuator bounds $\mathbf{u}_{\min},\, \mathbf{u}_{\max} \in \mathbb{R}^m$}
\Require{Armijo constant $c_1 \in (0,1)$, backtracking factor $\rho \in (0,1)$, initial step $\alpha_0 > 0$}
    \Comment{Added}

\vspace{0.5em}
\State{Define $J(\mathbf{u}) \triangleq \tfrac{1}{2}\,\mathbf{e}^{\top} W\,\mathbf{e} + \tfrac{1}{2}\,\mathbf{u}^{\top} R\,\mathbf{u}$,\; with $\mathbf{e} = \boldsymbol{\Theta}\!\left(\mathbf{x}_t,\mathbf{u}\right)\boldsymbol{\Xi} - \dot{\mathbf{x}}_{\mathrm{des},t}$}
    \Comment{Cost minimized by the search (added)}

\vspace{0.3em}
\State{\textbf{Initialization} $k \gets 0$; \quad $\mathbf{u}^{(0)} \gets \mathbf{u}_{\mathrm{act}}$}
    \Comment{Warm start}

\vspace{0.3em}
\State{$\dot{\mathbf{x}}_{\mathrm{pred}}^{(0)} \gets \boldsymbol{\Theta}\!\left(\mathbf{x}_t,\,\mathbf{u}^{(0)}\right)\boldsymbol{\Xi}$}
    \Comment{Evaluate with identified dynamics}
\State{$\mathbf{e}^{(0)} \gets \dot{\mathbf{x}}_{\mathrm{pred}}^{(0)} - \dot{\mathbf{x}}_{\mathrm{des},t}$}
    \Comment{Error}

\vspace{0.3em}
\While{$\left\|\mathbf{e}^{(k)}\right\|_2 > \varepsilon$}
    \Comment{Iterate until error is within tolerance}
    \vspace{0.3em}
    \State $\mathbf{J}^{(k)} \gets \dfrac{\partial\,\boldsymbol{\Theta}\!\left(\mathbf{x}_t,\,\mathbf{u}^{(k)}\right)}{\partial\,\mathbf{u}}\,\boldsymbol{\Xi} \in \mathbb{R}^{n \times m}$
        \Comment{Jacobian}

    \vspace{0.3em}
    \State{$\mathbf{g}^{(k)} \gets \left(\mathbf{J}^{(k)}\right)^{\!\top} W\,\mathbf{e}^{(k)} + R\,\mathbf{u}^{(k)}$}
        \Comment{Gradient of $J$}

    \vspace{0.3em}
    \State{$\alpha \gets \alpha_0$}
        \Comment{Begin Armijo backtracking (replaces fixed step)}
    \State{$\tilde{\mathbf{u}} \gets \mathrm{clip}\!\left(\mathbf{u}^{(k)} - \alpha\,\mathbf{g}^{(k)},\, \mathbf{u}_{\min},\, \mathbf{u}_{\max}\right)$}
        \Comment{Projected trial point}
    \While{$J\!\left(\tilde{\mathbf{u}}\right) > J\!\left(\mathbf{u}^{(k)}\right) + c_1\,\left(\mathbf{g}^{(k)}\right)^{\!\top}\!\left(\tilde{\mathbf{u}} - \mathbf{u}^{(k)}\right)$}
        \Comment{Sufficient-decrease test}
        \State{$\alpha \gets \rho\,\alpha$}
            \Comment{Shrink step}
        \State{$\tilde{\mathbf{u}} \gets \mathrm{clip}\!\left(\mathbf{u}^{(k)} - \alpha\,\mathbf{g}^{(k)},\, \mathbf{u}_{\min},\, \mathbf{u}_{\max}\right)$}
    \EndWhile{}
    \State{$\mathbf{u}^{(k+1)} \gets \tilde{\mathbf{u}}$}
        \Comment{Accept projected step}

    \vspace{0.3em}
    \State{$\dot{\mathbf{x}}_{\mathrm{pred}}^{(k+1)} \gets \boldsymbol{\Theta}\!\left(\mathbf{x}_t,\,\mathbf{u}^{(k+1)}\right)\boldsymbol{\Xi}$}
        \Comment{Coincides with final line-search evaluation}
    \State{$\mathbf{e}^{(k+1)} \gets \dot{\mathbf{x}}_{\mathrm{pred}}^{(k+1)} - \dot{\mathbf{x}}_{\mathrm{des},t}$}

    \State{$k \gets k + 1$}
\EndWhile{}

\vspace{0.3em}
\Return{$\mathbf{u}_{\mathrm{cmd},t} \gets \mathbf{u}^{(k)}$,\quad $\dot{\mathbf{x}}_{\mathrm{pred},t} \gets \dot{\mathbf{x}}_{\mathrm{pred}}^{(k)}$}
    \Comment{Actuator command and predicted model output for residual monitoring}
\end{algorithmic}
\end{algorithm}

Algorithm~\ref{alg:pgd_ca} summarizes the per step nonlinear control allocation routine. At each control update, the flight computer provides the commanded acceleration $\dot{\mathbf{x}}_{\mathrm{des},t}$ and the allocator starts from a nominal actuator vector $\mathbf{u}_{nom}$, chosen as the previous applied command (or a trim value), which serves as a warm start for the iterative solve. After each update, the actuator vector is clipped to remain within physical bounds. The final iterate is returned as the allocated command $\mathbf{u}_t$ to be applied by the actuators.

\subsection{Failure Scenarios and Adaptability}
Algorithm~\ref{alg:sindy_adapt_gd_activation} describes the online coefficient adaptation procedure that maintains model accuracy when the plant dynamics change during flight. The adaptation operates on a sliding window of length $H$, continuously comparing the predicted angular accelerations from the identified model against the corresponding sensor measurements. At each time step, the predicted and measured derivatives are appended to their respective buffers, and the oldest entries are discarded. The prediction error matrix $\mathbf{E}$ is then formed over the window, and a scalar mismatch metric $e_H$ is computed as the norm of $\mathbf{E}$.

\begin{algorithm}[htbp]
\caption{Online Coefficient Adaptation}\label{alg:sindy_adapt_gd_activation}
\begin{algorithmic}[1]

\Require Nominal coefficient matrix $\boldsymbol{\Xi}_0$

\vspace{0.5em}
\State \textbf{Initialization} $\boldsymbol{\Xi} \gets \boldsymbol{\Xi}_0$; \quad $\dot{\mathbf{X}}^{\mathrm{pred}}_H \in \mathbb{R}^{n \times H}$; \quad $\dot{\mathbf{X}}^{\mathrm{meas}}_H \in \mathbb{R}^{n \times H}$
    \Comment{Offline identified coefficients and buffers}

\vspace{0.3em}
\For{each control cycle $k$}

    \vspace{0.3em}
    \State $\dot{\mathbf{x}}_{\mathrm{pred},k}$ 
        \Comment{predicted model output from allocation}
    \State $\dot{\mathbf{x}}^{\mathrm{meas}}_{k}$
        \Comment{from sensors}
    \State Append $\dot{\mathbf{x}}_{\mathrm{pred},k}$ to $\dot{\mathbf{X}}^{\mathrm{pred}}_H$, drop oldest column
    \State Append $\dot{\mathbf{x}}^{\mathrm{meas}}_{k}$ to $\dot{\mathbf{X}}^{\mathrm{meas}}_H$, drop oldest column

    \vspace{0.3em}
    \State $\mathbf{E} \gets \dot{\mathbf{X}}^{\mathrm{pred}}_H - \dot{\mathbf{X}}^{\mathrm{meas}}_H$
        \Comment{Prediction error over the window}
    \State $e_H \gets \dfrac{1}{H}\|\mathbf{E}\|$
        \Comment{Mean mismatch metric}

    \vspace{0.3em}
    \noindent\hdashrule{\linewidth}{0.4pt}{2pt}

    \vspace{0.3em}
    \If{$e_H > \varepsilon_{\mathrm{threshold}}$}
        \Comment{trigger parameter update}
        \State $\mathbf{G} \gets \dfrac{1}{H}\,\mathbf{E}\,\boldsymbol{\Theta}^\top$
            \Comment{Gradient}
        \State $\boldsymbol{\Xi} \gets \boldsymbol{\Xi} - \eta\,\mathbf{G}$
            \Comment{Update coefficients}
    \EndIf
\EndFor
\vspace{0.3em}
\State \Return $\boldsymbol{\Xi}$
    \Comment{adapted coefficient to Algorithm~\ref{alg:pgd_ca}}
\end{algorithmic}
\end{algorithm}

The adaptation is activation gated: the coefficient matrix is updated only when the mismatch metric exceeds a predefined threshold $\varepsilon_{\mathrm{act}}$. This design avoids unnecessary parameter drift during nominal operation, where the offline identified model already provides sufficient accuracy. When the threshold is exceeded, indicating a meaningful discrepancy between the model and the true plant, a gradient step is computed to correct the coefficients. Because the identified model $\dot{\mathbf{X}} = \boldsymbol{\Theta}(\mathbf{x}, \mathbf{u})\,\boldsymbol{\Xi}$ is linear in the coefficient matrix $\boldsymbol{\Xi}$, the gradient of the prediction error with respect to $\boldsymbol{\Xi}$ reduces to a direct product of the error matrix and the library evaluations, $\mathbf{G} = \frac{1}{H}\,\mathbf{E}\,\boldsymbol{\Theta}_H^\top$, without requiring iterative solver. The coefficient matrix is then corrected in the direction that reduces the observed mismatch.

\subsection{Dynamic control allocation via MPC with identified dynamics}

Conventional allocators do not account for actuator dynamics, decoupling the
allocation decision from the physical response of the actuators~\cite{harkegaard2004dynamic}. In highly agile aircraft, this separation can be limiting: actuator bandwidth and rate limits directly shape the achievable response, and the allocator may generate commands that are feasible in a static sense but produce delayed or oscillatory realized deflections once actuator dynamics are applied.

Dynamic control allocation addresses these issues by solving for actuator commands over a short horizon while explicitly propagating actuator states and constraints. This formulation tends to produce smoother, more stable control inputs because it penalizes aggressive command sequences, anticipates actuator lag, and distributes the control effort in a way that remains feasible not only instantaneously but throughout the predicted transient. A closely related dynamic control allocation formulation that explicitly incorporates actuator dynamics can be found in~\cite{luo2004model}, where
MPC is used to optimize actuator commands over a prediction horizon while
accounting for nonnegligible actuator dynamics and hard constraints.

In this work, the predictive model used within the dynamic allocation problem is obtained from the identified analytical control effectiveness mapping. The mapping is evaluated online to predict the body rate accelerations for candidate actuator trajectories, and its analytical Jacobians are supplied to the nonlinear solver to accelerate convergence and improve numerical robustness. Actuator deflections are treated as dynamic states, and the allocator optimizes a sequence of commands to minimize the mismatch between predicted and commanded accelerations while regularizing control usage. 

In the dynamic allocation formulation, the MPC computes a sequence of commanded deflections $\mathbf{u}_{\mathrm{cmd},k}\in\mathbb{R}^{7}$ while explicitly propagating the realized deflections $\mathbf{u}_{\mathrm{act},k}\in\mathbb{R}^{7}$ through a first order actuator model. This distinction is central to the formulation: the identified analytical control effectiveness mapping is evaluated using the realized deflections, not the commanded ones, so that the predicted body rate response reflects the physical surface positions that the actuators can actually achieve within their bandwidth and rate limits. At each sampling instant, the flight computer provides the desired angular acceleration $\dot{\mathbf{x}}_{\mathrm{des}}\in\mathbb{R}^{3}$, and MPC optimizes the finite horizon command sequence
\(
\mathbf{u}_{\mathrm{cmd}}=\{\mathbf{u}_{\mathrm{cmd},k}\}_{k=0}^{N-1}
\)
to track this commanded acceleration while simultaneously regulating control effort and penalizing abrupt command changes that would stress actuator bandwidth:

\begin{align}
\underset{\mathbf{u}_{\mathrm{cmd}}}{\min} \; J &= \sum_{k=0}^{N-1}\Big[
(\dot{\mathbf{x}}_{\mathrm{pred},k}-\dot{\mathbf{x}}_{\mathrm{des},k})^\top \mathbf{Q}\, (\dot{\mathbf{x}}_{\mathrm{pred},k}-\dot{\mathbf{x}}_{\mathrm{des},k})  
\nonumber\\
&\quad +
(\mathbf{u}_{\mathrm{cmd},k}-\mathbf{u}_{\mathrm{cmd},k}^{\mathrm{nom}})^\top \mathbf{R}\, (\mathbf{u}_{\mathrm{cmd},k}-\mathbf{u}_{\mathrm{cmd},k}^{\mathrm{nom}})
\nonumber\\
&\quad +
(\Delta \mathbf{u}_{\mathrm{cmd},k})^\top \mathbf{S}\,(\Delta \mathbf{u}_{\mathrm{cmd},k})
\Big] \label{eq:nmpc-cost-aug} 
\end{align}

\begin{align}
\text{s.t.}\quad
\dot{\mathbf{x}}_{\mathrm{pred},k} &= \boldsymbol{\Theta}(\mathbf{x}_k,\,\mathbf{u}_{\mathrm{act},k})\,\boldsymbol{\Xi}, \label{eq:mpc-dyn}\\
    \dot{\mathbf{u}}_{\mathrm{act}} &= \frac{1}{\tau}\bigl(\mathbf{u}_{\mathrm{cmd}} - \mathbf{u}_{\mathrm{act}}\bigr), \label{eq:mpc-act}\\
\mathbf{u}_{\min} &\le \mathbf{u}_{\mathrm{cmd},k} \le \mathbf{u}_{\max}, \label{eq:mpc-pos} \\
\Delta \mathbf{u}_{\min} &\le \Delta \mathbf{u}_{\mathrm{cmd},k} \le \Delta \mathbf{u}_{\max}. \label{eq:mpc-rate} 
\end{align}

where $\dot{\mathbf{x}}_{\mathrm{pred},k} = \boldsymbol{\Theta}(\mathbf{x}_k, \mathbf{u}_{\mathrm{act},k})\,\boldsymbol{\Xi}$ is the body rate acceleration predicted by the identified mapping evaluated at the realized deflections, $\dot{\mathbf{x}}_{\mathrm{des},k}$ is the commanded acceleration from the flight computer held constant over the horizon, $\mathbf{Q}\succeq 0$ weights the acceleration tracking error, $\mathbf{R}\succ 0$ penalizes deviation of the commanded deflections from a nominal reference $\mathbf{u}_{\mathrm{cmd}}^{\mathrm{nom}}$, and $\mathbf{S}\succeq 0$ penalizes the slew $\Delta \mathbf{u}_{\mathrm{cmd},k} = \mathbf{u}_{\mathrm{cmd},k} - \mathbf{u}_{\mathrm{cmd},k-1}$ between successive commands. Together, the three weighting matrices allow the designer to shape how the allocator distributes effort across effectors and over time.

The constraints propagate the augmented dynamics using the identified analytical mapping together with the first order actuator model. Equation~\eqref{eq:mpc-act} propagates the realized deflections through first order actuator dynamics, modeled as where $\tau$ denotes the actuator time constant of each physical control surface. Together, constraints~\eqref{eq:mpc-pos} and~\eqref{eq:mpc-rate} enforce hard position and slew bounds on the commanded deflections, ensuring that the returned command sequence respects the physical actuator limits by construction.

\section{Results}\label{sec:results}
The results section is structured to evaluate the proposed allocator in the way it will be used in practice: as a closed loop module that must generate feasible actuator commands over extended, highly dynamic maneuvers. Rather than reporting only instantaneous, one step allocation errors, we assess full trajectory behavior because small modeling or allocation mismatches in nonlinear systems can accumulate over time and alter the subsequent state evolution. This is particularly important for data driven models, whose performance is inherently tied to the distribution of the training data. A learned control effectiveness mapping is optimized to match measured derivatives on the training set and is typically validated on a test set drawn from a similar distribution; however, in closed loop flight, the allocator actively shapes the trajectory, and small errors in the mapping can steer the system toward previously unseen regions of the state and input space. Once the trajectory drifts away from the known distributions, prediction errors can increase sharply and compound through integration, which may lead to  different behavior than indicated by pointwise error metrics. Consequently, the primary goal of this section is to demonstrate that the proposed method maintains accurate tracking and stable behavior over complete aggressive maneuvers, thereby validating both the generalization of the identified dynamics beyond the training distribution and the effectiveness of the nonlinear allocation routine under closed loop operation.

To place the proposed approach in the context of control allocation, we compare against both linear and nonlinear onboard models, which are widely used baselines in the literature and span different fidelity complexity tradeoffs. Linear onboard models remain the dominant baseline in aerospace applications due to their simplicity and speed, but their validity is tied to a local operating point and degrades under strong nonlinearities. High fidelity nonlinear onboard models capture these effects more accurately, yet they substantially increase computational cost. These two baselines therefore bracket the central tradeoff targeted by the proposed method: achieving near nonlinear model tracking performance with reduced model complexity and evaluation cost.

\subsection{Baseline Control Allocation Methodologies}
To benchmark our methodology against established practice, we follow the reference described by Durham~\cite{durham2017aircraft}. In this setup, the allocator receives the commanded angular acceleration and computes actuator commands that realize the demanded response. Two baseline variants are considered, differing in the fidelity of the onboard model used.

\subsubsection{Linear onboard model version.}
The model is linearized about a nominal flight condition and trim effectors $u_{\text{nom}}$, yielding the matrices $A$ and $B$ from the ADMIRE linearization tool~\cite{Admire_report}. The commanded actuator vector is computed as
\begin{align}
u_{\text{cmd}} &= u_{\text{nom}} + B^{-1}\!\left(\dot{x}_{\text{des}}-\dot{x}_{\text{nom}}\right) \\
\dot{x}_{\text{nom}} &= A x + B u_{\text{nom}}
\end{align}

\subsubsection{Nonlinear onboard model version.}
The nonlinear onboard model replaces the linearized prediction $\dot{x}_{\text{nom}}$ with nonlinear dynamics including aerodynamic data. The command law retains the same structure,
\begin{align}
u_{\text{cmd}} &= u_{\text{nom}} + B^{-1}\!\left(\dot{x}_{\text{des}}-\dot{x}_{\text{nom}}\right) \\
\dot{x}_{\text{nom}} &= f_{admire}(x,u_{\text{nom}})
\end{align}
where $f_{admire}(\cdot)$ denotes the nonlinear ADMIRE dynamics. When the nominal point is chosen as the current effector deflections, the nominal acceleration is simply the measured acceleration,
\begin{align}
\dot{x}_{\text{nom}} = \dot{x} = f(x,u)
\end{align}

For the nonlinear onboard model, the effectiveness matrix $B$ is obtained by finite differences method, perturbing one effector about its nominal value and recording the change in the acceleration outputs to form the corresponding column~\cite{durham2017aircraft}.

\begin{figure*}[ht]
    \centering
    \includegraphics[width=0.75\linewidth]{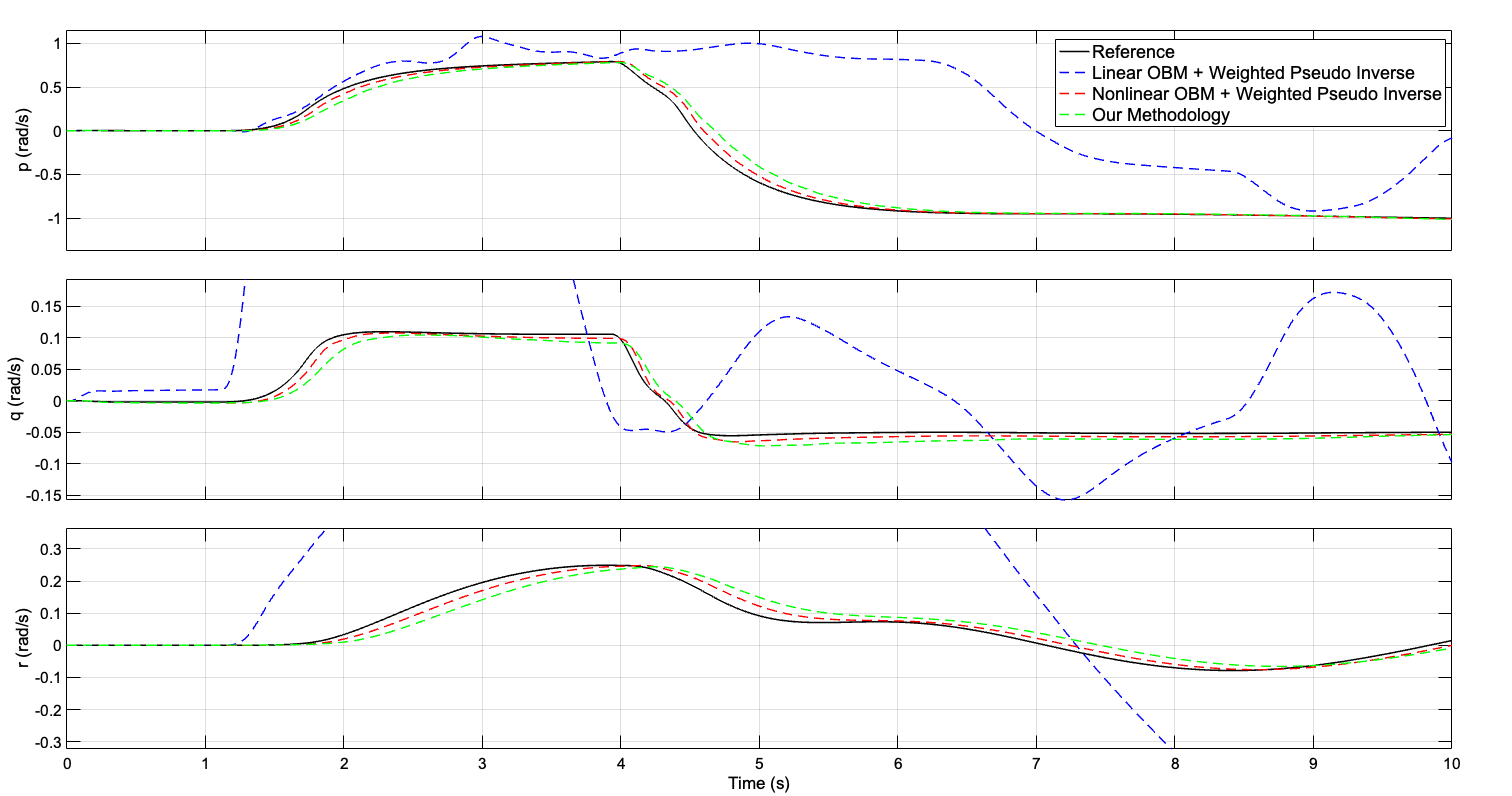}
    \caption{Tracking performance for a representative agile maneuver. Blue: linear onboard model baseline with weighted pseudo inverse allocation. Red: nonlinear onboard model baseline with weighted pseudo inverse allocation. Green: proposed methodology using the identified analytical control effectiveness mapping with gradient descent allocation.}\label{fig:resultsInversion_pqr_tracking}
\end{figure*}

\begin{figure*}[ht]
    \centering
    \includegraphics[width=0.75\linewidth]{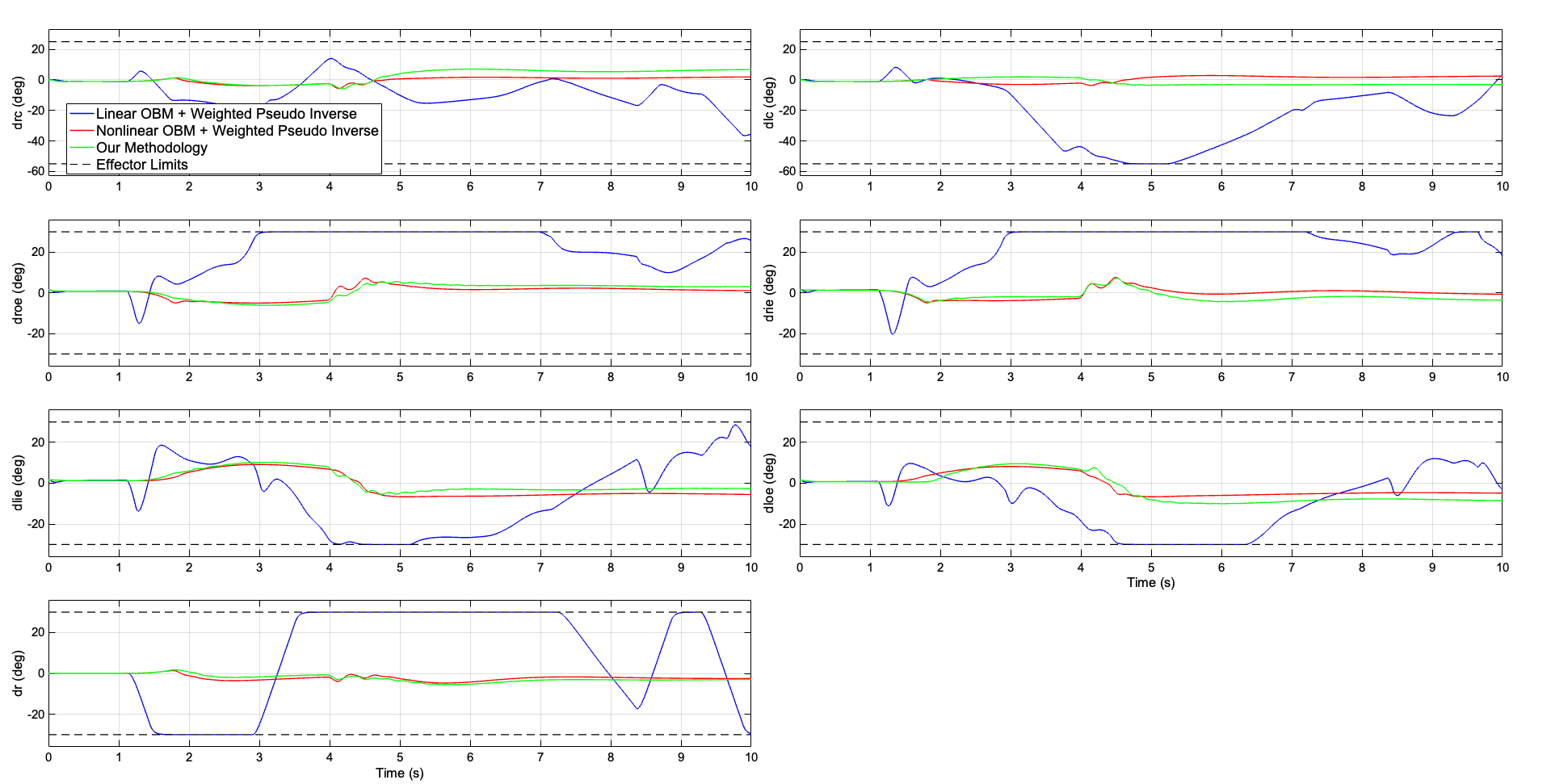}
    \caption{Control surface deflections during the representative maneuver in Fig.~\ref{fig:resultsInversion_pqr_tracking}. The linear onboard-model baseline generates large, oscillatory, and nonsmooth deflections as tracking degrades, whereas the nonlinear onboard model and the proposed methodology produce smooth and coordinated deflection histories.}\label{fig:resultsInversion_control_inputs}
\end{figure*}

\subsection{Control allocation with gradient descent}

All compared allocation schemes exhibit similar behavior in near trim, level flight conditions, where local linear effectiveness assumptions remain adequate and each method produces feasible actuator commands that follow the commanded response. In contrast, under highly agile maneuvers nonlinear effects become dominant and the linear onboard model allocator is no longer able to realize the flight computer inputs. To highlight this, the maneuver in Fig.~\ref{fig:resultsInversion_pqr_tracking} applies simultaneous roll, pitch, and yaw demands that drive large, coordinated deflections across the effectors. Under these conditions, the linear baseline shows pronounced tracking degradation and ultimately fails to follow the commanded response, while both the nonlinear onboard model baseline and the proposed method track the commands smoothly throughout the maneuver. Consistent with the tracking degradation in Fig.~\ref{fig:resultsInversion_pqr_tracking}, the corresponding actuator commands in Fig.~\ref{fig:resultsInversion_control_inputs} show that the linear onboard model baseline drives the effectors into large, oscillatory, and nonsmooth deflections as it attempts to compensate for the modeling mismatch. In contrast, both the nonlinear onboard model baseline and the proposed methodology produce smooth, coordinated deflection histories over the maneuver, indicating a more consistent realization of the commanded accelerations.

\begin{table*}[htbp]
\centering
\caption{Tracking error comparison over $N=100$ Monte Carlo trials}\label{tab:error_comparison}
\setlength{\tabcolsep}{10pt}
\renewcommand{\arraystretch}{1.4}
\begin{tabular}{l ccc}
\toprule
\textbf{Axis} 
    & \textbf{Linear OBM + WPI} 
    & \textbf{Nonlinear OBM + WPI} 
    & \textbf{Our Methodology} \\
    & $\mathrm{RMSE} \pm \sigma\ [\si{rad/s}]$
    & $\mathrm{RMSE} \pm \sigma\ [\si{rad/s}]$
    & $\mathrm{RMSE} \pm \sigma\ [\si{rad/s}]$ \\
\midrule
$p$ & $0.1338 \pm 0.1293$ & $0.0403 \pm 0.0384$ & $0.0824 \pm 0.0782$ \\
$q$ & $0.1091 \pm 0.1025$ & $0.0057 \pm 0.0028$ & $0.0135 \pm 0.0077$ \\
$r$ & $0.1220 \pm 0.1220$ & $0.0187 \pm 0.0186$ & $0.0356 \pm 0.0354$ \\
\bottomrule
\end{tabular}
\end{table*}

The proposed method achieves comparable closed loop behavior with nonlinear onboard model. This is confirmed quantitatively in Table~\ref{tab:error_comparison}, which reports the tracking RMSE over 100 Monte Carlo trials: the proposed method consistently outperforms the linear baseline across all axes and approaches the nonlinear onboard model accuracy, indicating that the identified model captures the control effectiveness variations needed for allocation.

\begin{figure*}
    \centering
    \includegraphics[width=0.75\linewidth]{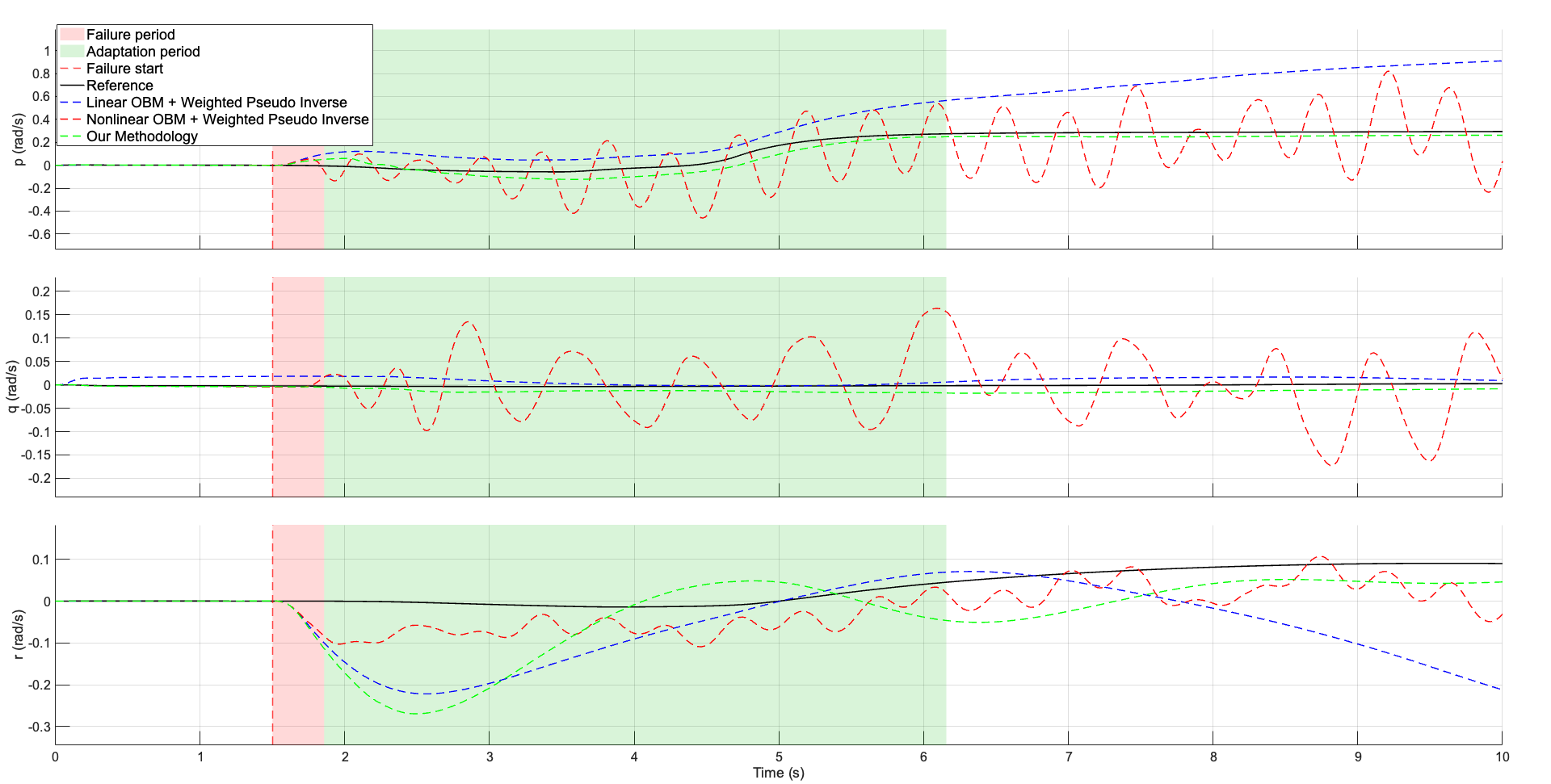}
    \caption{Tracking performance for a representative failure scenario. Blue: linear onboard model baseline with weighted pseudo inverse allocation. Red: nonlinear onboard model baseline with weighted pseudo inverse allocation. Green: proposed methodology using the identified analytical control effectiveness mapping with online adaptation. The green shaded region indicates the interval during which Algorithm~\ref{alg:sindy_adapt_gd_activation} is actively updating the coefficient matrix in response to the detected mismatch. The results show that the proposed methodology recovers accurate reference tracking after a short transient, achieving performance comparable to the nonlinear onboard model, while the linear baseline exhibits degraded tracking as maneuver aggressiveness increases.}\label{fig:results_failure_pqr}
\end{figure*}

\begin{figure*}
    \centering
    \includegraphics[width=0.75\linewidth]{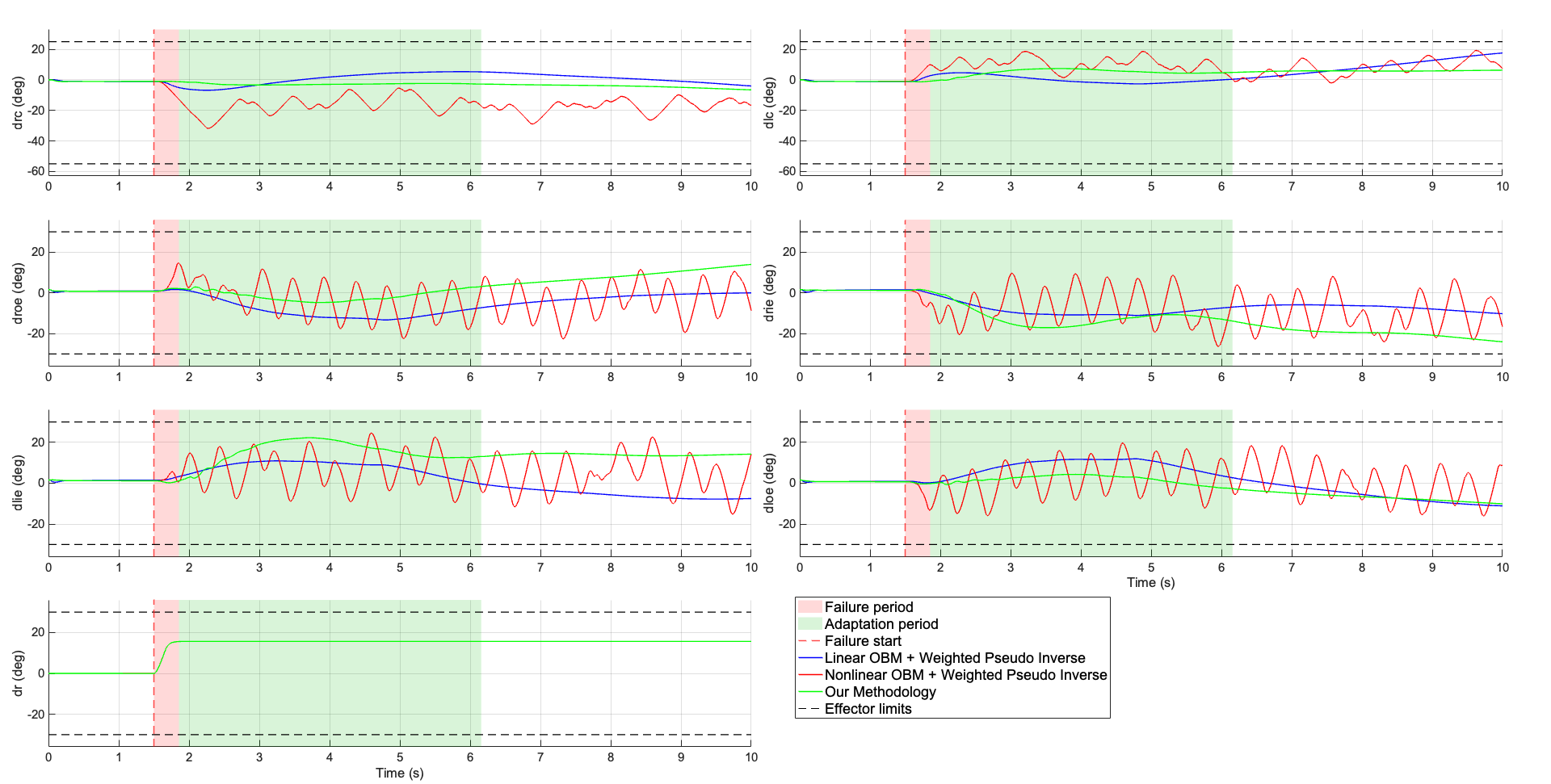 }
\caption{Control surface deflections in the actuator failure scenario. The failure is at $t=1.5\,\mathrm{s}$ by fixing the rudder at a constant positive deflection. After the prediction residual triggers the online adaptability mode, the proposed methodology redistributes the required control action across the remaining healthy effectors, which is reflected by increased and reshaped deflection commands and coincides with the tracking recovery in Fig.~\ref{fig:results_failure_pqr}.}\label{fig:results_failure_controls}
\end{figure*}

\subsection{Actuator failure adaptability}

To evaluate adaptability under off nominal conditions, we consider a representative failure scenario in which one aerodynamic effector becomes stuck at a fixed positive deflection. Specifically, the rudder is stuck at a constant offset while the remaining surfaces stay operational, introducing a persistent unmodeled change in control authority that requires the allocator to redistribute the demanded moments across the healthy effectors.

As shown in Fig.~\ref{fig:results_failure_pqr}, the actuator failure is introduced at $t=1.5\,\mathrm{s}$. Following the failure, the predicted accelerations from the identified mapping gradually diverge from the measured response, and once the prediction residual exceeds the adaptation threshold the method switches to its adaptation mode and updates the model parameters online. After a short transient, the tracking performance recovers, indicating that the updated mapping enables the allocator to re-distribute the control demand across the remaining effectors. This reconfiguration is also evident in the corresponding deflection histories in Fig.~\ref{fig:results_failure_controls}, where the allocator increases the usage of the healthy surfaces to compensate for the stuck rudder and maintain the commanded response.

\subsection{Actuator Aware Dynamic Control Allocation}

The previous subsection demonstrated that the identified analytical mapping achieves tracking accuracy comparable to the nonlinear onboard model when used inside a static, single step allocator. However, that formulation treats actuator commands as instantaneous decisions and does not account for the physical bandwidth and rate limits of the control surfaces. In practice, actuators are governed by actuator dynamics that introduce lag between the commanded and realized deflections; ignoring these effects can produce commands that are feasible in a static sense but result in delayed or oscillatory surface motion once actuator dynamics are applied. This subsection embeds the same identified mapping into a model predictive control formulation that explicitly propagates actuator states over a finite prediction horizon. To motivate this formulation, we note that conventional static allocation
methods perform well when actuators are sufficiently fast relative to the
control update rate. In the nominal ADMIRE configuration, all seven effectors
share the same time constant
$\boldsymbol{\tau}_{\mathrm{nom}}$, so the realized deflections closely
follow the commanded values within a single control cycle and the static
assumption is justified. To expose the effect of nonnegligible actuator
dynamics on allocation performance, we introduce a degraded configuration
$\boldsymbol{\tau}_{\mathrm{slow}}$ in which the four surfaces time constants
are increased to $0.50\;\si{s}$ while the canards and rudder retain their
nominal bandwidth as in Table~\ref{tab:actuator_time_constants}.
\begin{table}[htbp]
\centering
\caption{Actuator time constants for the nominal and degraded configurations.}\label{tab:actuator_time_constants}
\begin{tabular}{l S[table-format=1.2] S[table-format=1.2] S[table-format=1.2] S[table-format=1.2] S[table-format=1.2] S[table-format=1.2] S[table-format=1.2]}
\toprule
{Configuration} & {$\delta_{\mathrm{rc}}$} & {$\delta_{\mathrm{lc}}$} & {$\delta_{\mathrm{roe}}$} & {$\delta_{\mathrm{rie}}$} & {$\delta_{\mathrm{lie}}$} & {$\delta_{\mathrm{loe}}$} & {$\delta_{\mathrm{r}}$} \\
\midrule
$\boldsymbol{\tau}_{\mathrm{nom}}$ (\si{s})  & 0.05 & 0.05 & 0.05 & 0.05 & 0.05 & 0.05 & 0.05 \\
$\boldsymbol{\tau}_{\mathrm{slow}}$ (\si{s}) & 0.05 & 0.05 & 0.50 & 0.50 & 0.50 & 0.50 & 0.05 \\
\bottomrule
\end{tabular}
\end{table}

\begin{table*}[htbp]
  \centering
  \caption{Dynamic allocation comparison: Nonlinear OBM with weighted pseudo inverse and our methodology.}\label{tab:mpc_comparison}
  \sisetup{round-mode=places, round-precision=4}
  \begin{tabular}{@{}l S[table-format=1.4] S[table-format=1.4]@{}}
    \toprule
    {Tracking error} & {Nonlinear OBM + WPI} & {Our Methodology (Dynamic)} \\
    \midrule
    \quad RMSE $\dot{p}$ ($\mathrm{rad/s^2}$)  & 0.1366 & 0.1396 \\
    \quad RMSE $\dot{q}$ ($\mathrm{rad/s^2}$)  & 0.1737 & 0.1757 \\
    \quad RMSE $\dot{r}$ ($\mathrm{rad/s^2}$)  & 0.0401 & 0.0402 \\
    \quad RMSE mean         & 0.1168 & 0.1185 \\
\end{tabular}

  \vspace{1em}

  \sisetup{round-mode=places, round-precision=3}
  \begin{tabular}{@{}l
      S[table-format=1.3] S[table-format=1.3] S[table-format=1.3]
      S[table-format=1.3] S[table-format=1.3] S[table-format=1.3]@{}}
    \toprule
    & \multicolumn{3}{c}{Nonlinear OBM + WPI}
    & \multicolumn{3}{c}{Our Methodology (Dynamic)} \\
    \cmidrule(lr){2-4} \cmidrule(l){5-7}
    {Effector}
    & {Defl. ($\mathrm{rad}$)} & {Slew ($\mathrm{rad/s}$)} & {$U_{cmd} - U_{sensor}$ ($\mathrm{rad}$)}
    & {Defl. ($\mathrm{rad}$)} & {Slew ($\mathrm{rad/s}$)} & {$U_{cmd} - U_{sensor}$ ($\mathrm{rad}$)} \\
    \midrule
    $\delta_{\mathrm{rc}}$  & 0.017 & 0.142 & 0.006 & 0.090 & 0.318 & 0.012 \\
    $\delta_{\mathrm{lc}}$  & 0.018 & 0.163 & 0.006 & 0.037 & 0.063 & 0.002 \\
    $\delta_{\mathrm{roe}}$ & 0.027 & 0.185 & 0.014 & 0.016 & 0.039 & 0.003 \\
    $\delta_{\mathrm{rie}}$ & 0.033 & 0.238 & 0.016 & 0.020 & 0.334 & 0.007 \\
    $\delta_{\mathrm{lie}}$ & 0.029 & 0.155 & 0.015 & 0.020 & 0.115 & 0.004 \\
    $\delta_{\mathrm{loe}}$ & 0.025 & 0.187 & 0.014 & 0.023 & 0.323 & 0.007 \\
    $\delta_{\mathrm{r}}$   & 0.007 & 0.026 & 0.001 & 0.011 & 0.048 & 0.002 \\
    \bottomrule
  \end{tabular}
\end{table*}

\begin{figure*}
    \centering
    \includegraphics[width=0.75\linewidth]{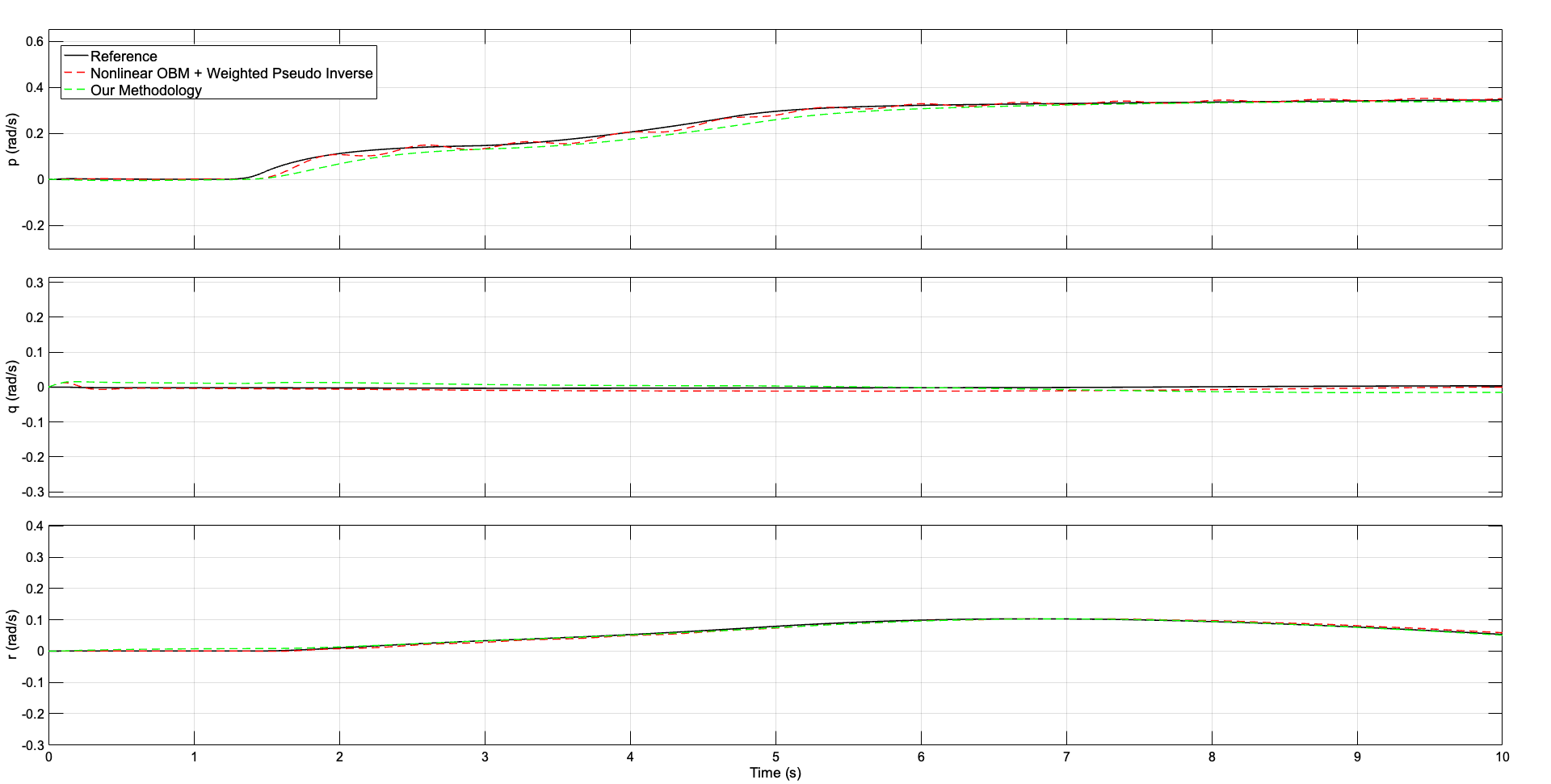}
    \caption{Tracking performance for a representative maneuver using dynamic allocation. Red: nonlinear onboard model baseline with weighted pseudo-inverse allocation. Green: proposed methodology using the identified analytical control effectiveness mapping embedded in the dynamic formulation.}\label{fig:results_mpc_pqr}
\end{figure*}

\begin{figure*}
    \centering
    \includegraphics[width=0.75\linewidth]{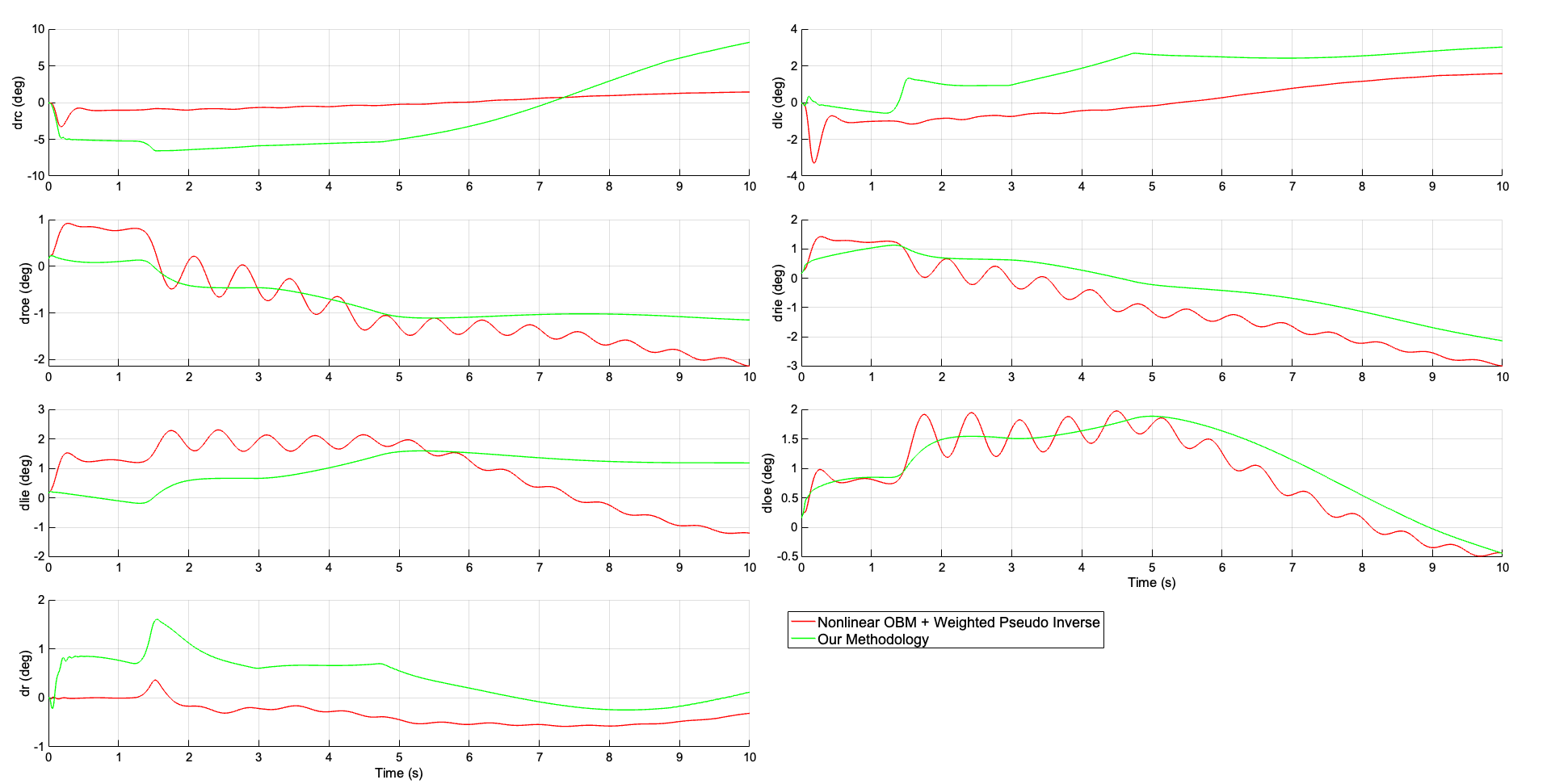}
    \caption{Control surface deflection histories for the representative maneuver in Fig.~\ref{fig:results_mpc_pqr}. Despite achieving comparable tracking to the baseline, the dynamic formulation produces smoother and more coordinated deflection commands across all effectors. This improvement results from the explicit actuator dynamics model.}\label{fig:results_mpc_controls}
\end{figure*}

In terms of tracking accuracy, the proposed methodology achieves performance comparable to the full nonlinear onboard model baseline, as shown in Fig.~\ref{fig:results_mpc_pqr}, confirming that the identified analytical mapping retains sufficient fidelity to support dynamic allocation over the prediction horizon. The deflection histories in Fig.~\ref{fig:results_mpc_controls} confirm that the static allocation baseline produces visibly oscillatory deflection commands on the slow elevon channels ($\delta_{\mathrm{roe}}$, $\delta_{\mathrm{rie}}$, $\delta_{\mathrm{lie}}$, $\delta_{\mathrm{loe}}$), as the allocator repeatedly demands deflections cannot be realized.  This is reflected quantitatively in Table~\ref{tab:mpc_comparison}: while both methods achieve comparable tracking RMSE across all body rate axes, the dynamic formulation reduces the commanded to realized deflection gap ($U_{\mathrm{cmd}}$ to $U_{\mathrm{act}}$) compared to the static baseline, confirming that the optimizer distributes control effort in a way that respects the physical actuators.

\subsection{Execution Time Comparison}
Table~\ref{tab:exec_times} summarizes the computational cost of the proposed 
methodology and the baseline implementations. For all onboard model based 
methods, the total computation time is decomposed into the model evaluation time 
and the allocation solver time, since both contributions are executed sequentially 
at each control update. The nonlinear onboard model evaluation cost is common 
across all nonlinear allocation methods and is therefore reported once. 

The results in Table~\ref{tab:exec_times} show that the proposed methodology 
achieves the lowest total execution time across all compared methods, including 
the linear onboard model baseline. This is primarily because the identified  dynamics and their analytical Jacobians are evaluated directly within 
the allocation solver, eliminating the separate model evaluation step and the 
need for finite difference perturbations. Among the nonlinear onboard model based 
methods, execution times vary considerably depending on the solver, with 
computationally demanding approaches such as linear 
programming incurring substantially higher costs than simpler pseudo inverse 
variants. These measurements confirm that the proposed approach delivers near 
nonlinear onboard model tracking accuracy while achieving the low computational 
cost across all evaluated methods.

\begin{table*}[htbp]
\centering
\caption{Execution time comparison of control allocation approaches.}\label{tab:exec_times}
\setlength{\tabcolsep}{10pt}
\renewcommand{\arraystretch}{1.3}
\begin{tabular}{l S[table-format=1.3] S[table-format=1.3] S[table-format=1.3]}
\toprule
\textbf{Method} 
    & \textbf{Model Eval.\ (s)} 
    & \textbf{Solver (s)} 
    & \textbf{Total (s)} \\
\midrule
Our Methodology (with gradient descent)                          & {--}  & {--}  & 0.161 \\
\midrule
\multicolumn{4}{l}{\textit{Nonlinear OBM + Control Allocation}} \\[2pt]
\quad Ganged into 3 pseudo effectors                 & 0.525 & 0.058 & 0.583 \\
\quad Weighted Pseudo Inverse              & 0.525 & 0.069 & 0.594 \\
\quad Direct Allocation                              & 0.525 & 1.781 & 2.306 \\
\quad Cascading Generalized Inverse                  & 0.525 & 0.125 & 0.650 \\
\quad Linear Programming (dual branch, $1$-norm)     & 0.525 & 5.780 & 6.305 \\
\midrule
\multicolumn{4}{l}{\textit{Linear OBM + Control Allocation}} \\[2pt]
\quad Weighted Pseudo 1Inverse                        & 0.150 & 0.045 & 0.195 \\
\bottomrule
\end{tabular}
\end{table*}

\section{Conclusion}\label{sec:conclusion}
This paper presented an integrated methodology for nonlinear control allocation in overactuated aerospace systems. It combines physics-informed sparse identification of the control effectiveness mapping with allocation solvers that exploit the resulting analytical Jacobians for fast, real-time computation, together with an online adaptation mechanism for graceful reconfiguration under changing flight conditions.

The mapping was identified with constrained SINDy and SR3 regression, embedding the known rigid-body rotational structure into the candidate library to narrow the search space relative to generic polynomial expansions; combined with an ensemble over independently identified models, this yielded a parsimonious, physically consistent model for an agile aircraft. The model was embedded in two complementary solvers: a projected gradient formulation that exploits the closed-form Jacobian, and a nonlinear MPC formulation that additionally propagates actuator dynamics over a finite horizon to anticipate bandwidth limitations and penalize command slew.

Monte Carlo evaluation confirmed tracking accuracy comparable to the nonlinear onboard-model baseline at a fraction of the computational cost. The online adaptation mechanism recovered tracking after an actuator failure by redistributing effort across healthy effectors, and the MPC variant eliminated the oscillatory deflections observed with static allocation under reduced actuator bandwidth.

Several directions remain. The study considered a single failure mode, a surface stuck at fixed deflection, and broader cases such as partial effectiveness loss should be assessed. Closed-loop stability under the online coefficient updates, and for the gradient-based formulation more generally, remains open, as does the practical tuning burden of the MPC variant, whose solver time, though reduced by the embedded explicit analytical model, stays substantially above the baselines. Robustness under realistic measurement noise also warrants dedicated study, since noise corrupts the derivative estimates used in both the offline identification and the online adaptation. Finally, because the methodology provides an explicit, differentiable model of the full control effectiveness landscape, it opens the possibility of not merely solving the allocation but actively exploiting the nonlinear dynamics, steering actuator combinations toward regions where effector couplings amplify the achievable moment, an opportunity inherently invisible to linear formulations.







\section*{ACKNOWLEDGMENT}
This work was funded by the Deutsche Forschungsgemeinschaft
(DFG, German Research Foundation) - Project ID 498601949 - TRR
364 SynTrac

\bibliographystyle{IEEEtran}
\bibliography{main}

\end{document}